\documentclass[aps,10pt,pra,twocolumn,floatfix,amsmath,amssymb]{revtex4-1}
\usepackage{physics}
\usepackage{graphicx}
\usepackage{mathdots}
\usepackage{bm}
\usepackage{times}

\usepackage[usenames,dvipsnames]{xcolor}
\usepackage{hyperref}
\hypersetup{
	colorlinks,
	allcolors=NavyBlue,
	linktoc=all
}

\usepackage[capitalize]{cleveref}
\crefformat{section}{Sec.~#2#1#3}

\renewcommand{\Re}{\mathrm{Re}}
\renewcommand{\Im}{\mathrm{Im}}
\newcommand{\cc}{\mathcal C}
\newcommand{\G}{\mathcal G}	
\renewcommand*\O{\mathcal{O}} 
\DeclareMathOperator{\sinc}{sinc}
\newcommand*\dagg{^{\dagger}}
\newcommand*\eps{\varepsilon}
\newcommand{\D}{\mathcal{D}}
\newcommand*\dens{\hat{\rho}}  
\renewcommand{\vec}[1]{\bm{#1}}
\newcommand*\mat[1]{\begin{pmatrix}#1\end{pmatrix}} 

\begin{document}
\title{Floquet approach to bichromatically driven cavity optomechanical systems}
\author{Daniel Malz}
\author{Andreas Nunnenkamp}
\affiliation{Cavendish Laboratory, University of Cambridge, Cambridge CB3 0HE, United Kingdom}
\date{\today}

\begin{abstract}
We develop a Floquet approach to solve time-periodic quantum Langevin equations in steady state. We show that two-time correlation functions of system operators can be expanded in a Fourier series and that a generalized Wiener-Khinchin theorem relates the Fourier transform of their zeroth Fourier component to the measured spectrum. We apply our framework to bichromatically driven cavity optomechanical systems, a setting in which mechanical oscillators have recently been prepared in quantum-squeezed states. Our method provides an intuitive way to calculate the power spectral densities for time-periodic quantum Langevin equations in arbitrary rotating frames.
\end{abstract}

\maketitle

\section{Introduction}
In a recent breakthrough, quantum squeezing of a mechanical oscillator has been demonstrated experimentally~\cite{Wollman2015,Pirkkalainen2015,Lecocq2015}.
The method has been analyzed first in Ref.~\cite{Mari2009}, but its full potential was realized in Ref.~\cite{Kronwald2013}.
It involves a standard optomechanical setup, comprising an optical cavity coupled to a mechanical oscillator, where the cavity mode is subject to unequally strong driving on both upper and lower mechanical sidebands.
This results in a Hamiltonian and consequentially quantum Langevin equations that are explicitly periodic in time.
Solving those is more difficult than stationary ones, since in general solutions contain all multiples of the fundamental frequency.

In this article, we develop a simple, yet powerful approach to find the steady state
of the bichromatically driven optomechanical system based on Floquet theory.
In effect, all system operators are split up into Fourier components, which individually obey stationary quantum Langevin equations.
As a result, any two-time correlation function of system operators $C(\tau,t)={\langle\hat A(t+\tau)\hat B(t)\rangle}$ is periodic in time $t$ and can be expressed in Fourier components,
a property that carries over to its Fourier transform $S(\omega,t)$.
Although a typical measurement only returns its time average, i.e., the zeroth Fourier component of $S(\omega,t)$,
the rotating components may carry information, as is the case for dissipative squeezing~\cite{Wollman2015,Pirkkalainen2015,Lecocq2015,Kronwald2013}.

Within our framework, we derive analytical expressions for the mechanical and optical spectrum within the rotating-wave approximation (RWA) for general detunings.
With the expressions for the Fourier components of system operators we provide,
it is straightforward to construct the spectrum in an arbitrary rotating frame.
This enables us to understand dynamical effects that occur when the drives are not exactly on the sidebands,
for example, how squeezing generation can fail or fail to be detected.
We show that there is a special frame in which rotating components
become part of the stationary spectrum and can be directly observed.
The method also elucidates how information about the system can be extracted
through a second, bichromatically driven ``readout'' mode,
an approach used in the experiments reported in Ref.~\cite{Lecocq2015}.
Our framework will be useful for other explicitly time-periodic quantum Langevin equations
and provides an intuitive way to understand power spectral densities in arbitrary rotating frames.

The remainder of this article is organized as follows. In \cref{sec:Model} we describe the model and our framework, how to obtain the solution, and familiarize ourselves with the properties of spectrum Fourier components.
Section~\ref{sec:RWA} exemplifies the technique through detailed analysis of dissipative squeezing.
This is followed by \cref{sec:readout},
which is concerned with the readout of the state of the mechanical oscillator through a second cavity mode.
Finally, we conclude in \cref{sec:conclusion}.

We note that Floquet theory has been developed on the level of the covariance matrix for an cavity optomechanical system with modulated coupling strength~\cite{Mari2009} as well as on the level of quantum master equations for numerical simulations of, for example, cavity quantum electrodynamics in Ref.~\cite{Papageorge2012}.

\section{Model}\label{sec:Model}
\begin{figure}
  \includegraphics[width=\linewidth]{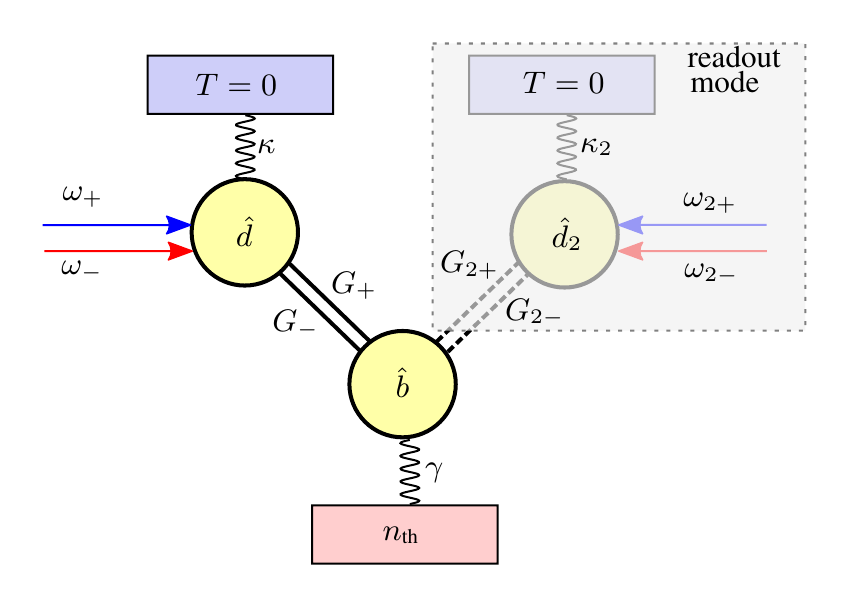}
  \caption{\textbf{Schematic of linearized quantum Langevin equations} \eqref{eq:langevin1} and \eqref{eq:langevin2}.
  The yellow circles depict harmonic oscillators,
  namely the mechanical mode with annihilation operator $\hat b$
  and two optical modes $\hat{d}$ and $\hat{d}_2$, respectively.
  The optical modes are coupled to the mechanical mode via radiation pressure (straight lines).
  Both optical modes are driven bichromatically, which leads to enhanced optomechanical coupling strengths $G_\pm$ and $G_{2\pm}$. The optical modes are also coupled to independent zero-temperature baths (blue) with rate $\kappa$ and $\kappa_2$, respectively.
  We will not consider the readout mode until \cref{sec:readout}.
  The mechanical mode is coupled to its own bath at a finite temperature (red) with a mean occupation $n_{\text{th}}$ and at a rate $\gamma$.}
  \label{fig:model}
\end{figure}
We consider a standard cavity optomechanical system in which the displacement of a mechanical oscillator modulates the frequency of an electromagnetic cavity mode.
For the most part we will consider one bichromatically driven cavity mode, but in \cref{sec:readout} we will include a second bichromatically driven cavity mode for readout. For a schematic, see \cref{fig:model}.

Without the second optical mode, the full Hamiltonian is
\begin{equation}
  H=H_{\text{sys}}+H_{\text{drive}}+H_{\text{baths}},
\end{equation}
where ($\hbar=1$)
\begin{subequations}
  \begin{align}
    H_{\text{sys}}&=\omega_{\text{cav}}a\dagg a+\Omega b\dagg b-g_0a\dagg a(b\dagg+b),\\
    H_{\text{drive}}&=(\alpha_+e^{-i\omega_+t}+\alpha_-e^{-i\omega_-t})a\dagg+\text{h.c.}
  \end{align}
\end{subequations}
$a,b$ are the bosonic annihilation operators of the cavity mode and the mechanical oscillator, respectively.
The cavity mode frequency is $\omega_{\text{cav}}$,
the mechanical frequency $\Omega$, the coupling strength via radiation pressure $g_0$,
and the driving strengths $\alpha_{\pm}$, which are associated with the drives with frequencies $\omega_\pm$.
A detailed derivation of the individual terms in this Hamiltonian can be
found for instance in Ref.~\cite{Aspelmeyer2014}.

To proceed, we split the light field into a coherent part and fluctuations,
move to a frame rotating with the frequency of the lower frequency laser,
$\hat a=e^{-i\omega_-t}(\bar a_-+\bar a_+e^{-i\delta t}+\hat d)$,
and linearize the Hamiltonian.
With the usual assumptions of Markovian baths, the resulting Hamiltonian
\begin{equation}
  H=-\Delta d\dagg d+\Omega b\dagg b-\left[ d\left( G_+e^{i\delta t}+G_- \right)(b\dagg+b)+\text{h.c.}\right]
  \label{eq:lin Hamiltonian}
\end{equation}
gives rise to Langevin equations~\cite{gardiner2004quantum,Clerk2010a} that are periodic in time
\begin{subequations}
  \begin{align}
  	\dot{d}&=\left( i\Delta-\frac{\kappa}{2} \right)d+\sqrt{\kappa}d_{\text{in}}
  	+i\left( G_+e^{-i\delta t}+G_-\right)(b\dagg+b),
  	\label{eq:langevin1}\\
  	\dot{b}&=\left(-i\Omega-\frac{\gamma}{2}\right)b+\sqrt{\gamma}b_{\text{in}}
  	+i\left[d\left(G_-+G_+e^{i\delta t}\right)+\text{h.c.}\right].
  	\label{eq:langevin2}
  \end{align}
\end{subequations}
Here, we have defined the enhanced optomechanical coupling constants $G_\pm=g_0\bar a_\pm$,
the detuning of the laser from the cavity mode $\Delta=\omega_--\omega_{\text{cav}}$,
and the difference between the two laser frequencies $\delta=\omega_+-\omega_-$.
Since we choose the frame of the lower frequency laser, $\delta>0$ always.
$b_{\text{in}},d_{\text{in}}$ are input noise operators with
$\langle d_{\text{in}}(t)d_{\text{in}}\dagg(t')\rangle=\delta(t-t')$,
$\langle d_{\text{in}}\dagg(t)d_{\text{in}}(t')\rangle=0$,
$\langle b_{\text{in}}(t)b_{\text{in}}\dagg(t')\rangle=(n_{\text{th}}+1)\delta(t-t')$, and
$\langle b_{\text{in}}\dagg(t)b_{\text{in}}(t')\rangle=n_{\text{th}}\delta(t-t')$.

\Cref{eq:langevin1,eq:langevin2} form the basis for our a\-na\-ly\-sis. 
We find their steady-state solution with a Floquet approach.

\subsection{Floquet Ansatz}
In order to solve \cref{eq:langevin1,eq:langevin2},
we express them in terms of Fourier components. We choose the conventions
\begin{subequations}
  	  \label{eq:fourier split}
	\begin{align}
  		d(t)&=\sum_{n=-\infty}^{\infty}e^{in\delta t}d^{(n)}(t),\\
  		d\dagg(t)&=\sum_{n=-\infty}^{\infty}e^{in\delta t}d^{(n)\dag}(t),
	\end{align}
\end{subequations}
and
\begin{subequations}
	\begin{align}
  		d^{(n)}    (\omega)&=\int_{-\infty}^\infty\dd{t}e^{i\omega t}d^{(n)}(t),\\
  		d^{(n)\dag}(\omega)&=\int_{-\infty}^\infty\dd{t}e^{i\omega t}d^{(n)\dag}(t).
	\end{align}
\end{subequations}
Note that these choices lead to $[d^{(n)}(\omega)]\dagg=d^{(-n)\dag}(-\omega)$.

The steady-state solution to \cref{eq:langevin1,eq:langevin2} is periodic~\cite{Yudin2016}
with period $2\pi/\delta$ and can be found by solving
\footnote{The Fourier components $\vec x^{(n)}(t)$ are not unique.
Given a solution $\{\vec x^{(n)}\}$, transformations such as
$\vec x^{(n)}(t)\to \vec x^{(n)}(t)+e^{ik\delta t}\vec y(t)$ and
$\vec x^{(n+k)}(t)\to\vec x^{(n+k)}(t)-\vec y(t)$ lead to other solutions.
However, these transformations leave the (physical) system operators
$\vec x(t)=\sum_ne^{in\delta t}\vec x^{(n)}(t)$ invariant
and we can show that the Fourier components of spectra are also unchanged.
In the main text we choose to put the noise operators entirely in the zeroth component equation.
The quantum Langevin equation is a first order ODE, which guarantees the uniqueness of its solution.
}
\begin{equation}
  i(\omega-\delta n)\vec x^{(n)}+\sum_{m=-\infty}^\infty A^{(m)}\vec x^{(n-m)}
  =-\delta_{n,0}\vec F_{\text{in}},
  \label{eq:floquet eom}
\end{equation}
where
\begin{equation}
  \begin{aligned}
  	\vec x^{(n)}&=\mat{d^{(n)}& b^{(n)}&d^{(n)\dag}&b^{(n)\dag}}^T,\\
  	\vec F_{\text{in}}&=\mat{\sqrt{\kappa}d_{\text{in}}&\sqrt{\gamma} b_{\text{in}}
	&\sqrt{\kappa}d\dagg_{\text{in}}&\sqrt{\gamma}b\dagg_{\text{in}}}^T,
  \end{aligned}
\end{equation}
and
\begin{subequations}
  \label{eq:A matrices}
	\begin{align}
  		A^{(0)}&=\mat{i\Delta-\frac{\kappa}{2}&iG_-&0&i\lambda G_-\\iG_-&-i\Omega-\frac{\gamma}{2}&i\lambda G_-&0\\
  		0&-i\lambda G_-&-i\Delta-\frac{\kappa}{2}&-iG_-\\-i\lambda G_-&0&-iG_-&i\Omega-\frac{\gamma}{2}},\\
  		A^{(-1)}&=iG_+
  		\left(\begin{array}{cc|cc}
  	  	  &\lambda&&1\\&&1&\\\hline\phantom{\lambda}&&&\\&&-\lambda&
  		\end{array}\right),
  		\label{eq:A-1}\\
  		A^{( 1)}&=iG_+\left(
  		\begin{array}{cc|cc}&&\phantom{-1}&\\\lambda&&&\\\hline&-1&&-\lambda\\-1&&&
  		\end{array}\right).
  		\label{eq:A+1}
	\end{align}
\end{subequations}
Here, we have introduced $\lambda$ to label the counterrotating terms.
In rotating-wave approximation (RWA) $\lambda=0$, else $\lambda=1$.

We can write \cref{eq:floquet eom} as an infinite-dimensional matrix
\begin{widetext}
  \begin{equation}
  	\mat{\ddots&\vdots&\vdots&\vdots&\iddots\\\cdots&i(\omega+\delta)+A^{(0)}&A^{(-1)}&A^{(-2)}&\cdots\\
  	\cdots&A^{(1)}&i\omega+A^{(0)}&A^{(-1)}&\cdots\\
  	\dots&A^{(2)}&A^{(1)}&i(\omega-\delta)+A^{(0)}&\cdots\\\iddots&\vdots&\vdots&\vdots&\ddots}
  	\mat{\vdots\\\vec x^{(-1)}\\\vec x^{(0)}\\\vec x^{(1)}\\\vdots}=\mat{\vdots\\0\\-\vec F_{\text{in}}\\0\\\vdots}.
  	\label{eq:infinite matrix}
  \end{equation}
\end{widetext}
In our case, only $A^{(0,\pm1)}$ are non-zero.
In the general case,
one has to truncate the infinite matrix~\eqref{eq:infinite matrix} to find an approximate solution.
In RWA the infinite set of equations decouples in sets of four, making the problem tractable analytically, see \cref{sec:RWA}.
\Cref{eq:infinite matrix} provides a visual tool for analyzing how the 4-by-4 blocks in each
entry are coupled to each other, which can be exploited to design new driving schemes.
For example, a block such as $A^{(n)}$ can be ``activated''
by either having an anharmonic drive with a nonzero $n$th Fourier component, 
or by adding a laser with frequency $\omega_-+n\delta$.
For details on how these matrices look like in general, see \cref{app:engineering}.

The advantage of splitting system operators up into Fourier components
is that these are governed by stationary quantum Langevin equations
and thus have time-independent expectation values and time-translation invariant correlation functions.
Therefore, any combination of Fourier components will have a well-defined spectrum
from which the measured spectra can be obtained in any rotating frame.

\subsection{Spectrum Fourier components}
One might ask which implications the time-periodicity of the quantum Langevin \cref{eq:langevin1,eq:langevin2} has on the properties of the measured spectra.
As has been alluded to above, the Fourier transform of the autocorrelator consists of Fourier components
and thus is not time-translation invariant.
In this section we introduce these Fourier components and mention some of their properties.
Finally, in a slight generalization of the Wiener-Khinchin (WK) theorem,
we show that the time-averaged power spectrum is the Fourier transform
of the zeroth Fourier component of the autocorrelator.

First, let us define
\begin{equation}
  S_{A\dagg A}(\omega,t)\equiv\int_{-\infty}^\infty\dd{\tau}e^{i\omega\tau}C_{AA}(\tau,t),
  \label{eq:spectrum}
\end{equation}
where $C_{AA}(\tau,t)=\ev{A\dagg(t+\tau)A(t)}$ is an autocorrelator.
We expect the steady state to be periodic, with period $2\pi/\delta$~\cite{Yudin2016}.
Therefore, $S_{A\dagg A}(\omega,t)$ can be expressed as a Fourier series
\begin{equation}
  S_{A\dagg A}(\omega,t)=\sum_{n=-\infty}^\infty e^{in\delta t}S_{A\dagg A}^{(n)}(\omega)
  \label{eq:spectrum fourier components}
\end{equation}
with Fourier components
\begin{equation}
  S_{A\dagg A}^{(m)}(\omega)=\sum_{n=-\infty}^\infty\int\frac{\dd{\omega'}}{2\pi}
  \ev{A^{(n)\dag}(\omega+n\delta)A^{(m-n)}(\omega')}.
  \label{eq:fourier components}
\end{equation}
By construction, the spectrum Fourier components encode all information about the autocorrelator $C_{AA}(\tau,t)$.
We will often refer to $S_{A\dagg A}(\omega,t)$ as ``spectrum'' although technically it is not a power spectrum in general. As we will show in \cref{app:stationaryWK}, in any given frame, the stationary part $S_{A\dagg A}^{(0)}$ is the physical power spectrum whereas other Fourier components $S_{A\dagg A}^{(m\not=0)}$ average out for long measurement times. This generalization of the WK theorem is consistent with the stationary case, 
where all Fourier components apart from the zeroth one vanish.
In one special rotating frame the rotating components become stationary
and can be directly measured, see \cref{sec:rot frame}.

Moreover, we can show that (proof in \cref{app:spectrum properties})
\begin{equation}
  \left[ S_{A\dagg B}^{(n)}(\omega) \right]^\dag=S_{B\dagg A}^{(-n)}(\omega+n\delta).
  \label{eq:reality}
\end{equation}
The stationary spectrum $S_{A\dagg A}^{(0)}(\omega)$ is thus real, but the other spectrum Fourier components are complex in general. 

Finally, we would like to mention that one can regard $S_{A\dagg A}(\omega,t)$
as a distribution of energy in time and frequency. 
Its marginal distributions are the stationary part
\begin{equation}
  S_{A\dagg A}^{(0)}(\omega)=\lim_{T\to\infty}\left[\frac{1}{T}\int_0^{T}\dd{t}S_{A\dagg A}(\omega,t)\right],
\end{equation}
and the variance as a function of time
\begin{equation}
  \ev{|A(t)|^2}=\int_{-\infty}^\infty\frac{\dd{\omega}}{2\pi}S_{A\dagg A}(\omega,t),
\end{equation}
both of which are guaranteed to be real and positive.

\subsection{The spectrum in a rotating frame}\label{sec:rot frame}
Although the rotating components of the spectrum drop out of the lab frame spectrum,
they can be observed in a special rotating frame.
In this section we show how rotating frames and spectra are expressed in our framework.

Let us start by defining a quadrature rotating at frequency $\nu$ and with an additional phase $\vartheta$
\footnote{
  We will use the notions ``measuring a rotating quadrature'' and ``measuring in a rotating frame'' interchangeably. Of course, all measurements will always be performed in a lab frame, but it can be more intuitive to think about rotating frames instead.
}
\begin{equation}
  \begin{aligned}
	X^\vartheta_\nu(t)&\equiv b(t)e^{i\nu t+i\vartheta}+b\dagg(t)e^{-i\nu t-i\vartheta}\\
	&=\sum_ne^{in\delta t}\left( b^{(n)}(t)e^{i\nu t+i\vartheta}+b^{(n)\dag}(t)e^{-i\nu t-i\vartheta}\right).
  \end{aligned}
  \label{eq:rotating quadrature}
\end{equation}

The autocorrelator of the rotating quadrature contains components rotating at $n\delta$ and $n\delta\pm2\nu$ in general
\begin{multline}
  S_{X^\vartheta_\nu X^\vartheta_\nu}(\omega,t)=\sum_{n,m}e^{i(n+m)\delta t}\\\times
  \bigg[ f_{b b}(n,m,\omega+n\delta+\nu)e^{2i(\nu t+\vartheta)}\\
	+f_{b\dagg b\dagg}(n,m,\omega+n\delta-\nu)e^{-2i(\nu t+\vartheta)}\\
	+f_{b b\dagg}(n,m,\omega+n\delta+\nu)\\
  +f_{b\dagg b}(n,m,\omega+n\delta-\nu)\bigg],
  \label{eq:rotating spectrum}
  \end{multline}
where we have introduced the shorthand
\begin{equation}
  f_{A\dagg B}(n,m,\omega)\equiv\int\dd{\tau}\exp(i\omega\tau)\ev{A^{(n)\dag}(t+\tau)B^{(m)}(t)}.
  \label{eq:shorthand}
\end{equation}
Note that the RHS of \cref{eq:shorthand} does \emph{not} depend on the  time $t$.
The Fourier components $A^{(n)},B^{(m)}$ are given by Langevin equations
without explicit time-dependence and thus their correlator is time-translation invariant.
Note that the sum $n+m$ tells us which lab frame
spectrum component $f(n,m,\omega)$ belongs to, as per~\cref{eq:fourier components}.

Equation~\eqref{eq:rotating spectrum} makes it clear that the case $\nu=\delta/2$ is special,
since in that case the terms $f_{b b}(n,-n-1,\omega+n\delta+\nu)$ 
and $f_{b\dagg b\dagg}(n,-n+1,\omega+n\delta-\nu)$ are part of the stationary spectrum.
We obtain
\begin{multline}
  S_{X^\vartheta_{\delta/2}X^\vartheta_{\delta/2}}^{(0)}(\omega)
  =S_{b b\dagg}^{(0)}(\omega+\delta/2)+S_{b\dagg b}^{(0)}(\omega-\delta/2)\\
  +\cos(2\vartheta)\left[S_{b b}^{(-1)}(\omega+\delta/2)+S_{b\dagg b\dagg}^{(1)}(\omega-\delta/2)\right].
  \label{eq:rot frame}
\end{multline}
It is real and positive. In particular, condition~\eqref{eq:reality} ensures that
$S_{b b}^{(-1)}(\omega+\delta/2)=[S_{b\dagg b\dagg}^{(1)}(\omega-\delta/2)]^*$.

The utility of these concepts will become clear in \cref{sec:variance in quadratures}
where we contrast spectra for dissipative squeezing in the lab frame
with those in the special rotating frame, see \cref{fig:spectrum fourier cpts}.

\section{Dissipative squeezing in the rotating-wave approximation}\label{sec:RWA}
In this section we derive analytic expressions for the system operator Fourier components, which enables a detailed study of dissipative squeezing and simultaneously serves to illustrate the advantages of our new framework.

To obtain an analytical solution, we will neglect counterrotating terms in \cref{eq:langevin1,eq:langevin2}, which results in
\begin{equation}
  \begin{aligned}
	\dot d&=\left( i\Delta-\frac{\kappa}{2} \right)d+\sqrt{\kappa}d_{\text{in}}
	+i\left( G_+e^{-i\delta t}b\dagg+G_-b \right),\\
	\dot b&=\left( -i\Omega-\frac{\gamma}{2}\right)b+\sqrt{\gamma}b_{\text{in}}
	+i\left( G_-d+G_+e^{-i\delta t}d\dagg\right).
  \end{aligned}
  \label{eq:RWA langevin}
\end{equation}
This is the rotating-wave approximation (RWA).
Note that by defining $\tilde d=e^{i\delta t/2}d$ and $\tilde b=e^{i\delta t/2}b$ it is possible to write
Eqs.~\eqref{eq:RWA langevin} in a frame where they become stationary.

Within RWA ($\lambda=0$) the infinite set of equations~\eqref{eq:floquet eom} decouples into sets of four.
Equivalently, we can make \cref{eq:infinite matrix} block-diagonal through a rearrangement of rows.
The blocks disconnected from input operators will decay and vanish in the steady state.
Thus, only two blocks (mutually hermitian conjugates) will contribute.
The problem reduces to solving
\begin{multline}
  \mat{\chi_c^{-1}(\omega)&-iG_-&0&-iG_+\\
  -iG_-&\chi_m^{-1}(\omega)&-iG_+&0\\
  0&iG_+&\chi_c^{-1*}(-\omega+\delta)&iG_-\\
  iG_+&0&iG_-&\chi_m^{-1*}(-\omega+\delta)}\\\times
  \mat{d^{(0)}(\omega)\\b^{(0)}(\omega)\\d^{(1)\dag}(\omega)\\b^{(1)\dag}(\omega)}
  =\mat{\sqrt{\kappa}d_{\text{in}}(\omega)\\\sqrt{\gamma}b_{\text{in}}(\omega)\\0\\0},
\end{multline}
with the cavity and mechanical response functions $\chi_c^{-1}(\omega)=\kappa/2-i(\omega+\Delta)$ and $\chi_m^{-1}(\omega)=\gamma/2-i(\omega-\Omega)$, respectively.

Inverting the matrix on the left-hand side, we can write the system operators in terms of input operators
\begin{equation}
  \mat{b^{(0)}(\omega)\\b^{(1)\dag}(\omega)}=
  \mat{a(\omega)&c(\omega)\\f(\omega)&g(\omega)}\mat{b_{\text{in}}(\omega)\\d_{\text{in}}(\omega)}.
  \label{eq:acfg mat}
\end{equation}
Analytic expressions for the auxiliary functions can be found in \cref{app:full solution}. Much of the physics can be understood by separating weak-coupling and strong-coupling effects, which we will discuss in turns below.

\subsection{Weak-coupling approximation}\label{sec:weak}
We can gain more insight when the coupling $G_\pm$ is small, such that second-order perturbation theory captures the main effects.

If $\Delta=-\Omega$ and writing $\delta=2\Omega+\eps$, we obtain to second order in $G_\pm$ (see \cref{app:weak})
\begin{subequations}
	\begin{align}
		\dot b^{(0)}&=\left( -i\tilde\Omega-\frac{\tilde\gamma}{2}\right)b^{(0)}
		+\frac{2iG_-}{\sqrt{\kappa}}d_{\text{in}}+\sqrt{\gamma}b_{\text{in}},\label{eq:weakeom1}\\
		\dot b^{(1)\dag}&=\left( i\tilde\Omega-i\delta-\frac{\tilde\gamma}{2} \right)b^{(1)\dag}
		-\frac{2iG_+}{\sqrt{\kappa}}d_{\text{in}},\label{eq:weakeom2}
	\end{align}
\end{subequations}
where
\begin{subequations}
	\begin{align}
	\tilde\gamma&=\gamma+\frac{4}{\kappa}\left( G_-^2-\frac{G_+^2}{1+4\eps^2/\kappa^2} \right),
  	  \label{eq:gammatilde}\\
	\tilde\Omega&=\Omega+\frac{G_+^2\eps}{(\kappa/2)^2+\eps^2}.
  	  \label{eq:Omegatilde}
	\end{align}
\end{subequations}

These equations provide several insights. First, in addition to the intrinsic mechanical damping,
$b^{(0)}$ is subject to ``optical damping''~\cite{Aspelmeyer2014}.
At $\eps=0$, this occurs with a rate $4\G^2/\kappa$, where $\G^2\equiv G_-^2-G_+^2$.
Since we are treating the problem in a frame where the red-detuned drive is stationary, 
it couples to the zeroth Fourier component with strength $G_-$.
Crucially, the optical input noise $d_{\text{in}}$ has opposite signs in the two equations.
The implications of that sign become clear if we consider the rotating quadrature (\ref{eq:rotating quadrature})
\begin{equation}
  X^{0}_{\nu}(t)=e^{i\nu t}\left[ b^{(0)}+e^{-i\delta t}b^{(-1)} \right]
  +e^{-i\nu t}\left[ b^{(0)\dag}+e^{i\delta t}b^{(1)\dag} \right].
  \label{eq:rwa rot quad}
\end{equation}
If $\delta=2\nu$, $b^{(0)}$ and $b^{(1)\dag}$ have the same phase factor
\begin{equation}
  X^0_{\delta/2}(t)=e^{i\delta t/2}\left[ b^{(0)}+b^{(1)\dag} \right]+\text{h.c},
\end{equation}
and Eq.~(\ref{eq:weakeom1}) gives
\begin{multline}
  \dot X^0_{\delta/2}=
  -\frac{\tilde\gamma}{2}X_{\delta/2}^{0}+\left( \frac{\delta}{2}-\tilde\Omega \right)X_{\delta/2}^{\pi/2}\\
  +\left\{e^{i\delta t/2}\left[ \frac{2i}{\sqrt{\kappa}}(G_--G_+)d_{\text{in}}
  +\sqrt{\gamma}b_{\text{in}}\right]+\text{h.c.}\right\}.
  \label{eq:rot quad eom}
\end{multline}

First, as is the case for all quadratures, the effective mechanical damping has an optical contribution.
Second, we see that in this particular rotating quadrature the optical noise is reduced,
which is also a feature of the exact equations of motion (see minus sign on RHS of \cref{eq:b solution} in \cref{app:full solution}),
and $X_{\delta/2}^0=X_-$ is the squeezed quadrature.
If $b^{(0)}$ and $b^{(1)\dag}$ do not have the same phase factor (for $\nu\neq\delta/2$),
then as time $t$ evolves, their relative phase changes,
such that sometimes the noises add and at other times they subtract, i.e.,
the quadrature we consider rotates relative to the squeezed and antisqueezed quadratures.
Third, note that the noises only subtract because both lasers are driving the same mode
and thus are subject to the same vacuum fluctuations.
If in addition $G_-=G_+$, this setup performs a quantum nondemolition (QND) measurement
of the rotating mechanical quadrature~\cite{Clerk2008}.
In \eqref{eq:rwa rot quad} we could set $\vartheta=\pi/2$, which would introduce a relative minus sign between the two square brackets, such that the noises add, to give the antisqueezed quadrature $X_+$.
Fourth, we note that the second term in \eqref{eq:rot quad eom} contains the conjugate quadrature. It is only non-zero if $\delta\neq2\Omega$.
Essentially, the mechanical quadratures naturally rotate at the mechanical frequency $\Omega$,
so the faster we rotate relative to $\Omega$ the quicker we will catch up with the quadrature $\pi/2$ ahead.
The resulting continuous mixing will play an important role in squeezing loss and heating,
cf.\ \cref{sec:squeezingloss,sec:heating}.

\begin{figure*}[t]
  \centering
  \includegraphics[width=\linewidth]{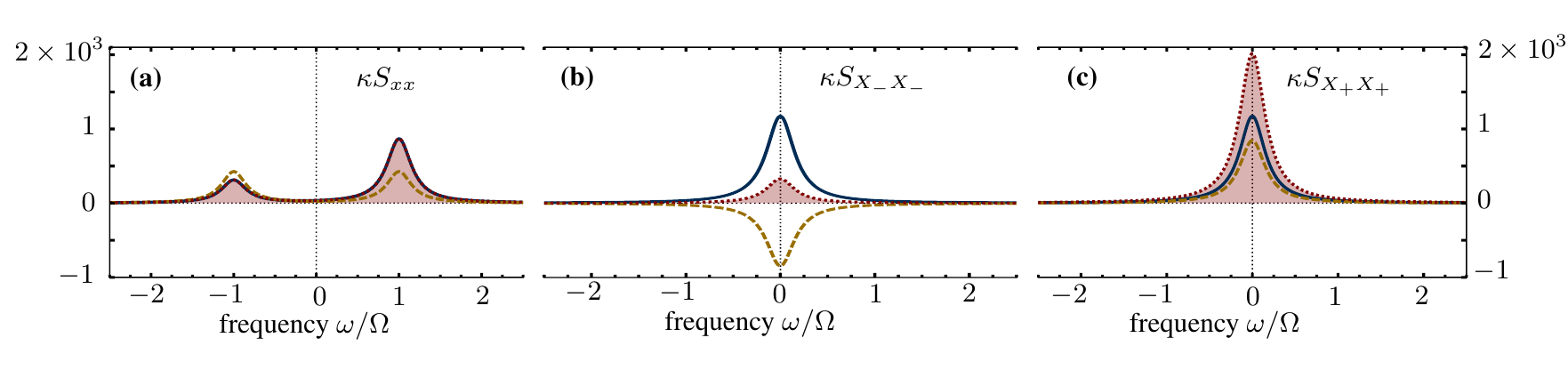}
  \caption{
  \textbf{Mechanical spectrum in the lab frame and the special rotating frame.} 
	\textbf{(a) Lab frame.} The stationary part $S_{xx}^{(0)}$ \eqref{eq:lab xx} is plotted in blue (solid) and has two peaks that stem from $S_{b\dagg b}^{(0)}$ (left) and $S_{b b\dagg}^{(0)}$ (right peak). $S_{xx}^{(0)}$ coincides with the measured spectrum
	for the position quadrature $x=b+b\dagg$ (red, filled). The yellow (dashed) curve is the absolute value of the sum of the rotating components $S_{b\dagg b\dagg}^{(1)}$ (left peak) and $S_{b b}^{(-1)}$ (right peak), and does not contribute to the lab frame 	spectrum.
	\textbf{(b), (c) Special rotating frame.} 
	The previously rotating spectrum components (still yellow and dashed) become stationary and thus part of the measured spectrum for the quadrature $X^\vartheta_{\delta/2}$ (red, dotted and filled), see \cref{eq:rot frame}.
	Their phase relation (encoded in $\vartheta$)
	determines whether they add to the stationary part (still blue and solid)
	to give the antisqueezed quadrature $X_+$ in (c), at $\vartheta=\pi/2$,
	or subtract from it to yield the squeezed quadrature $X_-$ in (b), at $\vartheta=0$.
	Parameters are $\gamma/\kappa=10^{-4},n_{\text{th}}=10,\cc=10^2,\Delta=-\Omega,\delta=2\Omega$.
	In RWA, the only effect of $\Omega/\kappa=.02$ is to determine the position of the peaks.
  }
  \label{fig:spectrum fourier cpts}
\end{figure*}

We Fourier transform~Eq.~(\ref{eq:weakeom1}) to obtain an approximation to 
\cref{eq:acfg mat}
\begin{equation}
  \begin{aligned}
	a(\omega)&\approx\sqrt{\gamma}\tilde\chi_m(\omega),\\
	c(\omega)&\approx2iG_-\tilde\chi_m(\omega)/\sqrt{\kappa},\\
	f(\omega)&\approx0,\\
	g(\omega)&\approx-2iG_+\tilde\chi_m^*(-\omega+\delta)/\sqrt{\kappa},
  \end{aligned}
  \label{eq:weak aux funs}
\end{equation}
where we have defined $\tilde\chi_m^{-1}(\omega)=\tilde\gamma/2-i(\omega-\tilde\Omega)$
and again have neglected terms $\O(G_\pm^3)$. For details see \cref{app:weak},
where we also write down an effective master equation that treats the cavity as an extra bath.

Using \cref{eq:weak aux funs,eq:fourier components} we write down the components that make up the mechanical spectrum for general detuning $\delta$
\begin{subequations}
	\begin{align}
		S_{b\dagg b}^{(0)}(\omega)&=|\tilde\chi_m(-\omega)|^2\left( \gamma n_{\text{th}}+\frac{4G_+^2}{\kappa} \right),\\
		S_{b b\dagg}^{(0)}(\omega)&=|\tilde\chi_m(\omega)|^2\left( \gamma(n_{\text{th}}+1)+\frac{4G_-^2}{\kappa}\right),\\
		S_{b b}^{(-1)}(\omega)&=-\frac{4G_-G_+}{\kappa}\tilde\chi_m(\omega)\tilde\chi_m(-\omega+\delta),\\
		S_{b\dagg b\dagg}^{(1)}(\omega)&=-\frac{4G_-G_+}{\kappa}\tilde\chi_m^*(\omega+\delta)\tilde\chi_m^*(-\omega).
	\end{align}
\end{subequations}
Integrating over the frequency $\omega$, we arrive at
\begin{subequations}
	\begin{align}
		\int_{-\infty}^{\infty}\frac{\dd{\omega}}{2\pi}S_{b\dagg b}^{(0)}(\omega)
		&=\frac{\gamma n_{\text{th}}+4G_+^2/\kappa}{\tilde \gamma},\\
		\int_{-\infty}^{\infty}\frac{\dd{\omega}}{2\pi}S_{b b\dagg}^{(0)}(\omega)
		&=\frac{\gamma(n_{\text{th}}+1)+4G_-^2/\kappa}{\tilde \gamma},\\
		\int_{-\infty}^{\infty}\frac{\dd{\omega}}{2\pi}S_{b b}^{(-1)}(\omega)
		&=-\frac{4G_-G_+/\kappa}{\tilde \gamma-i\eps},
  	  \label{eq:ad spec3}\\
		\int_{-\infty}^{\infty}\frac{\dd{\omega}}{2\pi}S_{b\dagg b\dagg}^{(1)}(\omega)
		&=-\frac{4G_-G_+/\kappa}{\tilde \gamma+i\eps},
  	  \label{eq:ad spec4}
	\end{align}
\end{subequations}
and we obtain the variance in the squeezed
and antisqueezed quadratures (which are rotating at the frequency $\delta/2$)
\begin{multline}
  \ev{X_\pm^2}=\frac{\gamma}{\tilde\gamma}(2n_{\text{th}}+1)+\frac{4}{\kappa\tilde\gamma}(G_+^2+G_-^2)\\
  \pm\frac{8G_-G_+}{\kappa\tilde\gamma}\left(\frac{1}{1+\eps^2/\tilde\gamma^2}\right),
  \label{eq:weak squeezing formula}
\end{multline}
where we have defined the detuning of the higher-frequency laser from the upper mechanical sideband as $\eps\equiv\delta-2\Omega$.
Term-by-term, the variance contains a reduced (if $\G^2>0$) occupancy due to the extra optical damping, a positive term due to the noise added by the drives,
and a term that can be negative due to the aforementioned noise canceling effect
of the two drives in one of the quadratures, see \cref{eq:weakeom1,eq:weakeom2}.
In the antisqueezed quadrature, the noises add.
The optically enhanced damping rate $\tilde\gamma$ reduces to the one for sideband cooling for $\eps\gtrsim\kappa$. In that limit the last term on the RHS of \cref{eq:weak squeezing formula} vanishes and the two quadratures have equal variances.
\Cref{eq:weak squeezing formula} is then very close to the expected result, apart from the extra noise term $4G_+^2/\kappa\tilde\gamma$, which at this level of approximation does not depend of the detuning $\eps$.

\subsection{Variance in the squeezed and antisqueezed quadratures}
\label{sec:variance in quadratures}
In \cref{sec:weak} we found that the quadrature in which the optical noises cancel most is the one rotating at half the laser frequency difference $\delta/2$. With the analytical solution at hand, we can go a more direct way and ask which  phase $\vartheta$ will have the smallest (or largest) quadrature variance. In agreement to what we found above, $\vartheta$ will have to depend on time with angular velocity $\delta/2$.

Let us consider a lab frame quadrature $X_{\nu=0}^{\vartheta}$, with variance
\begin{equation}
  \ev{(X_0^\vartheta)^2}=1+2\sum_ne^{in\delta t}\Xi_{bb}^{(n)}
  +2\Re\left[ e^{2i\vartheta}\sum_ne^{in\delta t}\Xi_{b\dagg b}^{(n)} \right],
  \label{eq:variance in fixed quadrature}
\end{equation}
where
\begin{equation}
  \Xi^{(n)}_{AB}\equiv \int S_{A\dagg B}^{(n)}(\omega) \frac{\dd{\omega}}{2\pi}.
  \label{eq:xi}
\end{equation}
Note that by \cref{eq:reality} the second term on the RHS of \cref{eq:variance in fixed quadrature}
is always real. The variance is minimal for
\begin{equation}
  \vartheta=\frac{\pi}{2}-\frac 12\arg\left[ \sum_ne^{in\delta t}\Xi_{b\dagg b}^{(n)} \right].
\end{equation}
In RWA, the only non-zero $\Xi_{b\dagg b}^{(n)}$ is the one with $n=-1$,
which turns out to be real and negative.
This results in $\vartheta(t)=\delta t/2$, the squeezed quadrature is rotating.
So, even though we started off not knowing that we would have to consider a rotating quadrature, 
the result emerged naturally.

We can calculate the maximum and minimum variance
\begin{equation}
  \ev{X_\pm^2}=1+2\sum_ne^{in\delta t}\Xi_{bb}^{(n)}
  \pm2\left|\sum_ne^{in\delta t}\Xi_{b\dagg b}^{(n)}\right|.
  \label{eq:direct variance}
\end{equation}
For the position quadrature $x=X_0^{\vartheta=0}$ and in RWA,
we obtain
\begin{equation}
  \ev{x(t)^2}=1+2\Xi_{bb}^{(0)}+2|\Xi_{b\dagg b}^{(-1)}|\cos(\delta t-\phi),
  \label{eq:position operator variance}
\end{equation}
where we have written the complex number $\Xi_{b\dagg b}^{(-1)}$
in terms of its absolute value and phase $\phi$
\footnote{
  The phase $\phi$ is primarily set by the relative phase of the lasers. 
  In terms of their intensity beating, the squeezed quadrature can be found at or near the maximum intensity.
  They do not coincide if $\delta\neq 2\Omega$,
  in which case the squeezed quadrature lags slightly behind.
  The assumption that the coherent amplitudes $\bar a_\pm$ are real leads to
  $\phi\approx\pi$ (equality if $\delta=2\Omega$).
}.
Note that \cref{eq:position operator variance} is the squared width in $x$-direction
of an ellipse with major and minor axis $\langle X^2_\pm\rangle^{1/2}$,
rotating at frequency $\delta/2$, with an initial tilt of $\phi/2$.
This is no coincidence---the Wigner density of a squeezed state is an ellipse.
There is one frame in which it is stationary, whereas in all other frames,
the ellipse is rotating, and thus a measurement of the variance will 
return an average over both quadratures.
Note that rotating the ellipse by $\pi$ maps it onto itself, so we can take $\vartheta\in[0,\pi)$.

The conclusion is that in order to detect the squeezing we have to follow the quadrature and make the measurement in a special rotating frame.
The necessity to ``follow'' the quadrature has been mentioned in the discussion of QND measurements in Ref.~\cite{Clerk2010a}.
The fact that we need to measure the rotating spectrum components to observe squeezing substantiates the claim that essential information can be hidden in rotating components of spectra.
In the literature, this special case is what characterizes a so-called ``phase-sensitive'' detector,
also called ``phase nonpreserving amplifier'' in Ref.~\cite{Clerk2010a}.
Such a detector requires an external ``clock'' (here the beating of the laser drives)
in order to keep track of the rotating quadrature, as noted in Ref.~\cite{Braginsky1980}.

In \cref{fig:spectrum fourier cpts} we illustrate how these concepts take form on the level of the mechanical spectra and plot the physical spectrum $S_{X^\vartheta_\nu X^\vartheta_\nu}^{(0)}(\omega)$ in the three most relevant cases. The first panel corresponds to $\nu=0=\vartheta$, i.e.~the spectrum of the lab frame position quadrature $X^0_0=x=b+b\dagg$.
  The left and right peak correspond to contributions of $\langle b\dagg b\rangle$
  and $\langle bb\dagg\rangle$, respectively.
  The absolute value of the rotating terms is shown as well. 
  In general, they are complex, with a phase depending on $t$ and $\vartheta$.

The second and third panel in \cref{fig:spectrum fourier cpts}
  are the spectra in the special rotating frame $\nu=\delta/2$.
  The first consequence of going into a rotating frame is that the peaks are displaced 
  (not unilaterally, because $b$ and $b\dagg$ get opposite phases, see \cref{eq:rotating spectrum}).
  In this frame, all peaks end up on top of each other.
  \Cref{eq:reality} ensures that the imaginary parts of the rotating Fourier components cancel.
  while their relative angle in the complex plane is $2\vartheta$.
  We show the two cases in which they (individually) are entirely real, $\vartheta=0,\pi/2$,
  and thus have the strongest effect.
  $\vartheta=0,\pi/2$ corresponds to the squeezed and antisqueezed quadrature $X_{\mp}$
  (second and third panel), with the smallest and largest variance, respectively.

  \subsection{Squeezing for exact sideband driving}\label{sec:exact squeezing}
  \begin{figure}[t]
  	\includegraphics[width=\linewidth]{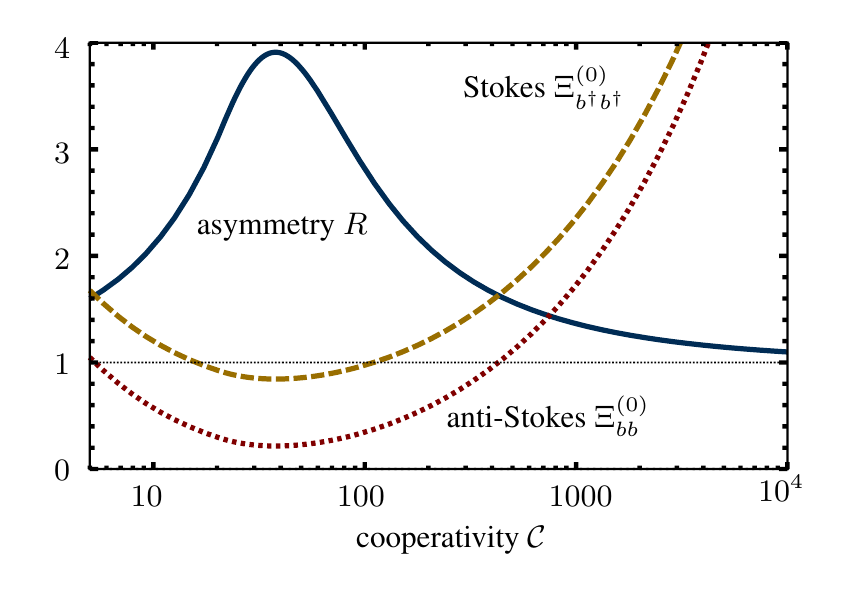}
	\caption{\textbf{Sideband asymmetry.} Weights of the left and the right peak of the mechanical spectrum $S_{xx}^{(0)}$	in the lab frame as a function of cooperativity $\cc$. 
	Left peak weight $\Xi_{bb}^{(0)}$ is labelled anti-Stokes (red dotted),
	right peak weight $\Xi_{b\dagg b\dagg}^{(0)}$ is labelled Stokes (yellow dashed).
	Blue (solid) is their ratio $R=\Xi_{b\dagg b\dagg}^{(0)}/\Xi_{bb}^{(0)}$.
	Parameters are $\gamma/\kappa=10^{-4},n_{\text{th}}=10,\Delta=-\Omega,\delta=2\Omega$.
	$\Omega/\kappa$ is irrelevant in RWA.
  }
  \label{fig:asymmetry}
\end{figure}
Reference~\cite{Kronwald2013} considered the case where the drives are on the sidebands, 
i.e., $\delta=2\Omega$ and $\Delta=-\Omega$.
Within RWA, the physical spectrum (cf.\ \cref{eq:rot frame}) of the squeezed quadrature in a frame rotating with the mechanical frequency $\Omega$ is given by
\begin{equation}
  S_{X^0_\Omega X^0_\Omega}(\omega)=\frac{\kappa|\chi_c(\omega+\Omega)|^2(G_--G_+)^2+\gamma(2n_{\text{th}}+1)}
  {|\chi_m^{-1}(\omega+\Omega)+\chi_c(\omega+\Omega)\G^2|^2}.
  \label{eq:X1spectrum}
\end{equation}
This is a roundabout way to arrive at the desired result,
as in this case it is easier to directly solve \cref{eq:langevin1,eq:langevin2} in a rotating frame, but our method is more general, enabling general detunings, rotating frames, and even beyond-RWA numerics.

Integrating \cref{eq:X1spectrum} over frequency, we obtain the variance of the squeezed and antisqueezed quadratures
\begin{multline}
  \ev{X_{\pm}^2}=\frac{1}{\kappa+\gamma}\left[(2n_{\text{th}}+1)\gamma
   	\left(1+\frac{\kappa}{\gamma+4\G^2/\kappa}\right)\right.\\+
  \left.\frac{4(G_-\pm G_+)^2}{\gamma+4\G^2/\kappa}\right],
  \label{eq:squeezing formula}
\end{multline}
where $X_-=X_{\nu=\Omega}^{\vartheta=0}$ and $X_+=X_{\nu=\Omega}^{\vartheta=\pi/2}$.
The result agrees with Ref.~\cite{Wollman2015},
where $(\kappa+\gamma)^{-1}\approx\kappa^{-1}$ was approximated.

Within our framework it is straightforward to find out how squeezing looks like in the lab frame. In \cref{fig:spectrum fourier cpts}(a) we plot the spectrum of the lab frame position operator $X^0_0=x=b+b\dagg$
\begin{multline}
  S_{xx}^{(0)}(\omega)=\frac{(n_{\text{th}}+1)\gamma+\kappa G_-^2|\chi_c(\omega)|^2}{|\chi_m^{-1}(\omega)+\chi_c(\omega)\G^2|^2}\\
  +\frac{\gamma n_{\text{th}}+\kappa G_+^2|\chi_c(-\omega)|^2}{|\chi_m^{-1}(-\omega)+\chi_c(-\omega)\G^2|^2}.
  \label{eq:lab xx}
\end{multline}
It has two peaks as long as we do not consider the strong-coupling regime, where normal-mode splitting occurs. We call them Stokes $(\omega=\Omega)$ and anti-Stokes $(\omega=-\Omega)$~\cite{Aspelmeyer2014}. As we have discussed, the squeezing terms are not present.

The weights of the left and right (anti-Stokes and Stokes) peak are the integrals
$\Xi_{bb}^{(0)}$ and $\Xi_{b\dagg b\dagg}^{(0)}$, respectively.
$\Xi$ is defined in \cref{eq:xi}.
The ratio of Stokes to anti-Stokes is the asymmetry $R=\Xi_{b\dagg b\dagg}^{(0)}/\Xi_{bb}^{(0)}$.
In \cref{fig:asymmetry} we plot the weights as a function of cooperativity $\cc$ for the ``optimal driving strength'' as defined in Ref.~\cite{Kronwald2013}
\begin{equation}
  G_-=\sqrt{\frac{\cc\kappa\gamma}{4}},\qquad G_+=G_-\left( 1-\sqrt{\frac{1+2n_{\text{th}}}{\cc}} \right).
  \label{eq:optimal driving strength}
\end{equation}
At low cooperativities, the asymmetry increases with cooperativity. Physically, this is because the system is cooled. However, as the coupling strength is increased further, the asymmetry decreases and approaches unity.
This is due to the fact that dissipative squeezing leads to a squeezed, thermal state
with an effective temperature that increases with the degree of squeezing.
In the lab frame, the squeezing terms are not a part of the spectrum,
so we expect that the quadrature variance and the weight of both peaks increase. This leads to a decrease in the asymmetry $R$ as a function of cooperativity $\mathcal{C}$.

\subsection{Squeezing loss due to detuning}\label{sec:squeezingloss}
Instead of having both drives exactly on the sidebands as in \cref{sec:exact squeezing},
in this section we will study the behavior of the system when the drives are detuned from the sidebands. Here, we will only analyze the case $\Delta=-\Omega,\delta=2\Omega+\eps$, i.e., the red drive remains on the sideband. Changing the detuning of the cooling drive will lead to an instability for $G_+>\gamma$.

\begin{figure}[t]
  \includegraphics[width=\linewidth]{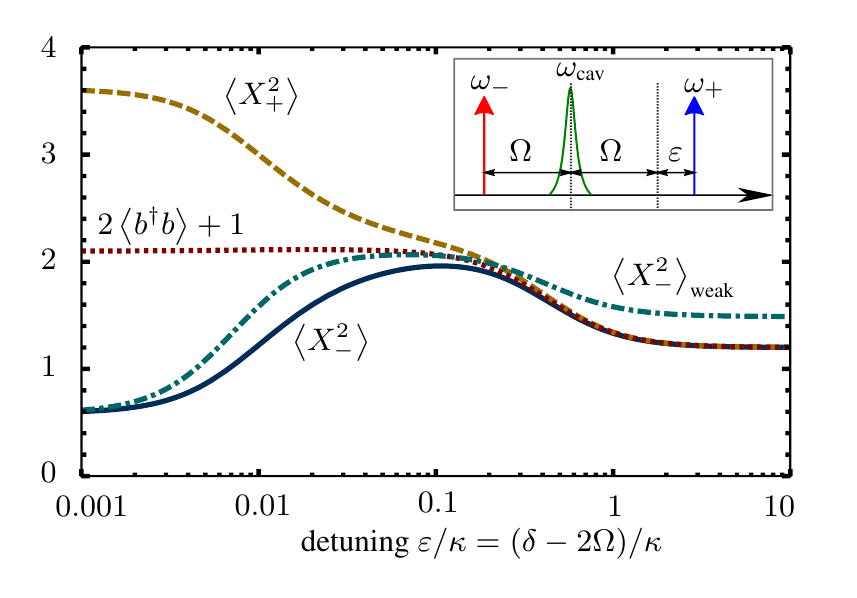}
  \caption{\textbf{Squeezing loss due to detuning.}
  Blue (solid) is the variance of the squeezed quadrature $\ev{X_-^2}$,
  orange (dashed) the antisqueezed quadrature $\ev{X_+^2}$, and red (dotted) is $2\ev{b\dagg b}+1$ as a function of detuning of the blue drive $\eps=\delta-2\Omega$.
  They have been obtained from \cref{eq:direct variance}.
  We also show the weak-coupling result \cref{eq:weak squeezing formula} for the squeezed quadrature in turquoise (dash-dotted).
  The inset shows the driving scheme with the two driving frequencies $\omega_\pm$ relative to 
  the cavity frequency $\omega_{\text{cav}}$.
  Parameters are $\gamma/\kappa=10^{-4},n_{\text{th}}=10,\cc=10^2,\Delta=-\Omega$, so that $\tilde\gamma\approx0.02\kappa$.
  $\Omega/\kappa$ is irrelevant in RWA.
}
\label{fig:squeezingloss}
\end{figure}
In \cref{fig:squeezingloss}, we plot the variance of the two quadratures, their average $2\langle b\dagg b\rangle+1$, and the weak-coupling result for the variance of the squeezed quadrature $\langle X_-\rangle$ as a function of the detuning $\eps$.
There are two scales on which effects occur~\footnote{
The lag of the squeezed quadrature behind the laser beating mentioned in \cref{sec:variance in quadratures}
  is negligible for the physics that we would like to discuss.
  In addition, we will assume that the time scale on which the measurement is performed is large 
  compared to any other time scale in the problem. 
  If that were not the case, we could observe rotating spectrum components that decay as $\sinc(\nu T)$
  for their respective frequency $\nu$ and measurement time $T$.
}.

The larger scale is the cavity mode dissipation rate $\kappa$. Detunings on this scale render the detuned drive ineffective such that only cooling remains. In particular, we see that the occupation and the variance of both quadratures decreases, as the influence of the blue drive becomes weaker. Note that by this point both quadrature variances are already almost equal.

The smaller scale is the effective mechanical damping $\tilde\gamma=\gamma+4\G^2/\kappa$, 
introduced in \cref{sec:weak}.
For $\eps\sim\tilde\gamma$, squeezing has disappeared and for strong driving an instability occurs, see \cref{sec:heating}. In \cref{fig:squeezingloss} the loss of squeezing is evidenced by the two quadrature variances becoming equal.
On this scale it does not matter whether we move the blue drive away or the red, 
as long as $\eps\ll\kappa$, as these effects are due to the mismatch between the beating frequency
of the two lasers $\delta$ and the mechanical frequency $\Omega$.
The beating can be thought of as a stroboscopic measurement of one of the quadratures every half period, akin to the scheme in Ref.~\cite{Vasilakis2015}.
For finite detuning $\eps$ the measured quadrature starts to rotate at frequency $\eps/2$
with respect to mechanical quadrature, so $2/\eps$ is the timescale on which the squeezed and antisqueezed quadratures mix and interchange, see Eq.~\eqref{eq:rot quad eom}.
In this sense, we are probing dynamical effects---they only become visible if their timescale
is comparable to $\eps^{-1}$.
The mixing eventually mitigates squeezing entirely
at $\eps\sim\tilde\gamma$, i.e., when the mixing rate balances the squeezing rate
as predicted in the weak-coupling approximation~\eqref{eq:weak squeezing formula}.
The weak-coupling approximation \eqref{eq:weak squeezing formula} does not correctly capture the sideband cooling limit, the noise added by the blue-detuned drive does not vanish in the limit $\eps\to\infty$, as discussed below \cref{eq:weak squeezing formula}.

\subsection{Heating and parametric instability}\label{sec:heating}
\begin{figure}[t] 
  \centering
  \includegraphics[width=\linewidth]{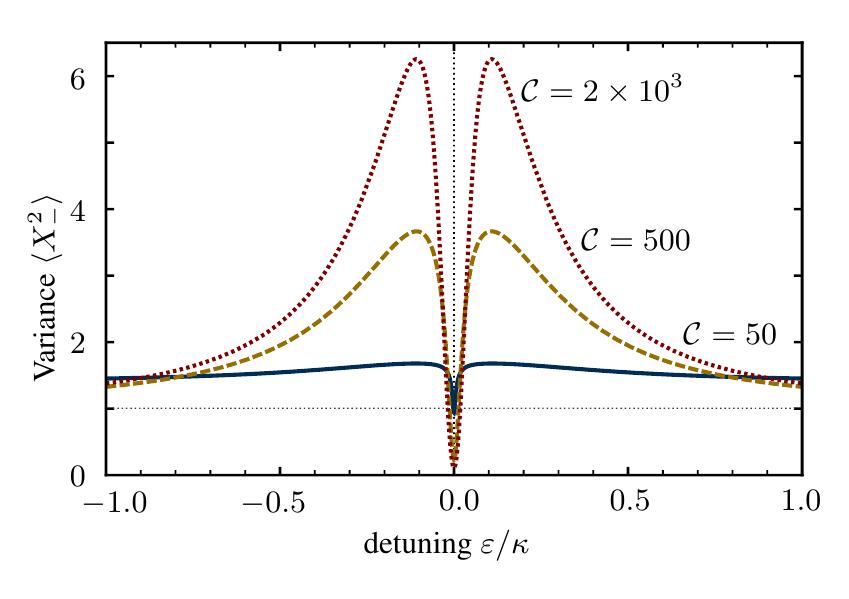}
  \caption{\textbf{Heating due to detuning.}
  Blue (solid), yellow (dashed), and red (dotted) are
  the squeezed quadrature variances $\langle X_-^2\rangle$ for cooperativities $\cc=50,500,2000$
  as a function of detuning $\eps/\kappa$.
  Here, $\gamma/\kappa=10^{-4},n_{\text{th}}=10,\Delta=-\Omega$. $\Omega/\kappa$ is irrelevant in RWA.}
  \label{fig:heating}
\end{figure}
We now turn to the strong-coupling effects.
If the system is coupled more strongly, with $\G$ approaching $\kappa$, the minimum variance of the squeezed quadrature saturates at the lower bound $\langle X_-^2\rangle\to\gamma(1+2n_{\text{th}})/(\kappa+\gamma)$, see \cref{eq:squeezing formula} or Ref.~\cite{Kronwald2013}.
In this regime, moving one of the lasers away from the sidebands, i.e., $\delta\neq2\Omega$, will result in a heating effect, and an instability for very strong coupling, see \cref{fig:heating,app:stability analysis}.

In \cref{fig:heating}, we plot the squeezed quadrature variance $\langle X_-^2\rangle$
as a function of the detuning of the blue laser $\eps$ for cooperativities $\cc=50,500,2000$.
As we couple more strongly, heating occurs in addition to squeezing loss.
From $\eps=0$, and for large enough $\cc$, the squeezed quadrature variance first increases steeply,
reaches a peak, and then decreases.
The peak corresponds to the point where the system is closest to instability, 
whereas the decay for $\eps\sim \kappa$ is the convergence to usual sideband cooling, as mentioned before.
The heating effect has been mentioned in Ref.~\cite{Pirkkalainen2015}
where it was used to tune the lasers to the mechanical sidebands.
Again, we find the separation of time scales: squeezing loss and heating for $\eps\sim\tilde\gamma$ and cooling for $\eps\sim\kappa$.
We analyze the instability further in \cref{app:stability analysis}.

\section{Measurement with second cavity mode}\label{sec:readout}
The ideas introduced above can be nicely illustrated if we study how the mechanical spectrum can be observed through a second, weakly coupled ``readout'' mode.
Our approach will be the same as above, with two lasers pumping a single cavity mode,
except that in this section the mechanical oscillator is a black-box with a fixed,
unknown spectrum that we would like to measure.
We will neglect the measurement backaction on the mechanical oscillator, 
an assumption that is excellent for QND measurements and reasonable for weak coupling.

Analogous to the first cavity mode $\hat{d}$, the linearized quantum Langevin equation for the second cavity mode $\hat{d}_2$ is
\begin{multline}
  \dot d_2=\left(i\Delta_2-\frac{\kappa_2}{2}\right)d_2+\sqrt{\kappa_2}d_{2\text{in}}\\
  +i\left(G_{2+}e^{-i\delta_2 t}+G_{2-}\right)(b\dagg+b),
  \label{eq:second mode langevin}
\end{multline}
where $\Delta_2=\omega_{2-}-\omega_{\text{cav},2}$ is the detuning of the lower frequency laser from the frequency of the second cavity mode,
$\delta_2=\omega_{2+}-\omega_{2-}$ is the frequency difference between the blue and the red drive on the second cavity mode,
$\kappa_2$ the dissipation rate of the second cavity mode,
and $G_{2\pm}$ are the enhanced optomechanical couplings, see \cref{fig:model}.

We can apply an analysis as above to find the most general spectra measured through the second cavity mode. 
For details, we refer to \cref{app:second mode}.
We split \cref{eq:second mode langevin} up into Fourier components of the two frequencies present
\begin{equation}
  d_2(t)=\sum_{n,m}e^{in\delta t+im\delta_2t}d_2^{(m,n)}(t).
\end{equation}
Generalized to two frequencies, the stationary spectrum is
\begin{multline}
  S_{d\dagg _2d_2}^{(0)}(\omega)\\=\sum_{n,m}\int\frac{\dd{\omega'}}{2\pi}
  \ev{d_2^{(n,m)\dag}(\omega+n\delta+m\delta_2)d_2^{(-n,-m)}(\omega')}.
\end{multline}

If $\delta\neq\delta_2$ (and are not multiples of each other),
$b$ does not have any components commensurate with $\delta_2$, and hence
\begin{equation}
  b^{(n,m)}=0,\quad\forall m\neq0.
  \label{eq:b fourier cpts are zero}
\end{equation}
The stationary part simplifies to
\begin{equation}
  S_{d\dagg _2d_2}^{(0)}(\omega)=|\chi_2(-\omega)|^2\left[G_{2-}^2S_{xx}^{(0)}(\omega)+G_{2+}^2S_{xx}^{(0)}(\omega+\delta_2)\right].
  \label{eq:d2 stationary part}
\end{equation}
$x=X^0_0$ here, as always, refers to the non-rotating position quadrature in the lab frame.
Therefore, the only effect of having a second drive is that now there are two copies of the mechanical spectrum superposed with a different weights and shifted by $\delta_2$ relative to each other.
Furthermore, both are filtered by the response function of the cavity mode
$\chi_2(\omega)=[\kappa_2/2-i(\omega+\Delta_2)]^{-1}$.
This case corresponds to the ``non-QND'' measurement in Ref.~\cite{Lecocq2015}.
It is an average over the squeezed and antisqueezed quadrature,
see \cref{sec:variance in quadratures,fig:spectrum fourier cpts}.

A special case is $\delta_2=\delta$, in which \cref{eq:b fourier cpts are zero} does not hold.
Instead, we find for the stationary part of the $d_2$ spectrum
\begin{multline}
  	  S_{d\dagg _2d_2}^{(0)}(\omega)=
  	  |\chi_2(-\omega)|^2\left\{ G_{2-}^2S_{xx}^{(0)}(\omega)+G_{2+}^2S_{xx}^{(0)}(\omega+\delta)\right.\\
  	  \left.+G_{2-}G_{2+}\left[S_{xx}^{(-1)}(\omega+\delta)+S_{xx}^{(1)}(\omega)\right]
	\right\}.
	\label{eq:special case d2 spectrum}
\end{multline}
Note that here the rotating parts of $S_{xx}$ contribute to $S_{d\dagg _2d_2}^{(0)}$.

In RWA, only $b^{(0)},b^{(0)\dag},b^{(-1)},b^{(1)\dag}$ are non-zero.
Depending on the cavity linewidth $\kappa_2$, the prefactor $|\chi_2(-\omega)|^2$
more or less sharply picks out the contribution at $\omega=-\Delta_2$.
This causes a suppression of counterrotating terms.
So, if we make the readout mode a good cavity with $\kappa_2\ll\Omega$ and choose $\Delta_2=-\delta/2$,
then we can make a second RWA (this time for the second optical mode) and we are left with
\begin{align}
	\begin{split}
  S_{d\dagg _2d_2}^{(0)}(\omega)&=|\chi_2(-\omega)|^2
  \left[G_{2-}^2S_{b\dagg b}^{(0)}(\omega)+G_{2+}^2S_{b b\dagg}^{(0)}(\omega+\delta)\right.\\
  &\left.\qquad+G_{2+}G_{2-}(S_{b b}^{(-1)}(\omega+\delta)+S_{b\dagg b\dagg}^{(1)}(\omega))\right]
  \end{split}\nonumber\\
  &=|\chi_2(-\omega)|^2G_2^2S_{X^0_{\delta/2}X^0_{\delta/2}}(\omega+\delta/2),
  \label{eq:d2 xx spectrum}
\end{align}
where in the last line we have chosen $G_{2+}=G_{2-}\equiv G_2$,
and identified the physical spectrum~\eqref{eq:rot frame}.
Thus, this is a measurement of a rotating quadrature.
In order to find out which terms contribute in \eqref{eq:d2 xx spectrum},
it is helpful to refer to the plot of spectrum Fourier components in RWA shown in \cref{fig:spectrum fourier cpts},
and remember that $|\chi_2(-\omega)|^2$ picks out contributions around $\omega=-\delta/2$.
If additionally $\delta=2\Omega$, this measurement is QND, as in Ref.~\cite{Lecocq2015}.

\section{Conclusion}\label{sec:conclusion}
In this article we presented a framework to deal with time-periodic quantum Langevin equations that builds on Floquet theory. Since the steady-state solution is periodic,
it amounts to splitting system operators up into their Fourier components \eqref{eq:fourier split}.
The spectrum Fourier components~\eqref{eq:spectrum fourier components} can be used to calculate power spectra in any rotating frame~\eqref{eq:rotating spectrum}.
This opens a new perspective to understand the relation between the measured spectra and rotating frames, as discussed in \cref{sec:rot frame}.

We exemplify the new tool by studying a bichromatically driven cavity optomechanical system
that has garnered a large amount of interest recently~\cite{Wollman2015,Pirkkalainen2015,Lecocq2015}.
This setting has been used to prepare a mechanical oscillator in a quantum-squeezed state, following the proposal~\cite{Kronwald2013}.
Using the full analytical solution in the rotating-wave approximation, 
we shed light on the squeezing mechanism and provide some intuition for the behavior
of bichromatically driven systems (\cref{sec:RWA}).

Looking ahead, the presented framework can be used to map time-periodic quantum Langevin equations to familiar, coupled, stationary ones, albeit---as usual for Floquet methods---infinitely many such equations.
Where an exact analytical solution is not feasible, an approximation can be found by
truncating the infinite matrix~\eqref{eq:infinite matrix}.
We would like to point out Ref.~\cite{Ranzani2015} as a graphical tool to approximate the
inverses of matrices such as \cref{eq:infinite matrix}, to any desired order in the coupling.
Furthermore, it may prove beneficial to identify conditions under which exact solutions can be found.

\section*{Acknowledgments}
We are grateful to Aashish Clerk, Florian Marquardt, Amir Safavi-Naeini, John Teufel, and, in particular, Tobias Kippenberg for stimulating and insightful discussions. A.N. holds a University Research Fellowship from the Royal Society and acknowledges additional support from the Winton Programme for the Physics of Sustainability. D.M. is supported by an EPSRC studentship.

\appendix

\section{Floquet engineering}\label{app:engineering}
In the case studied in the main text, the infinite matrix~\eqref{eq:infinite matrix}
only contains $A^{(0)},A^{(\pm1)}$, the others being zero.
We describe how to activate more blocks and their general structure below.

One can think of $A^{(0)}$ as the fundamental building block and of $A^{(\pm n)}$ for $n>0$ as
contributions that oscillate with $n\delta$ and therefore are capable of coupling fundamental blocks
a distance $n$ away from each other.

Any periodic driving with period $T=2\pi/\delta$, either due to anharmonic drives or several harmonic ones,
can be expressed as a Fourier series with fundamental frequency $\delta$.
Usually, the drive frequencies are offset by the cavity mode frequency and some detuning, i.e.,
\begin{equation}
  \omega_n=\omega_{\text{cav}}+\Delta+in\delta.
\end{equation}
It is useful to define the matrices, see Eqs.~(\ref{eq:A-1}) and (\ref{eq:A+1}),
\begin{equation}
  A_+\equiv
  \left(\begin{array}{cc|cc}&&\phantom{1}&\\1&&&\\\hline&-1&&-1\\-1&&&
  \end{array}\right),\,
  A_-\equiv\left(\begin{array}{cc|cc}
  	&1&&1\\&&1&\\\hline\phantom{1}&&&\\&&-1&
  \end{array}\right).
\end{equation}

If we assume a driving Hamiltonian of the form
\begin{equation}
  H_{\text{drive}}=e^{-i(\omega_{\text{cav}}+\Delta)t}\left( \sum_n\alpha_n e^{-in\delta t} \right)\hat{a}\dagg+\text{h.c.},
\end{equation}
we can linearize the Hamiltonian by a displacement operation like the one used in the main text, with
\begin{equation}
  \hat{a}=e^{-i(\omega_{\text{cav}}+\Delta)t}\left(\sum_n\bar a_n e^{-in\delta t}+\hat{d}\right).
\end{equation}
Defining $J_n=\bar a_n g$, the  enhanced optomechanical coupling strengths,
we can write
\begin{equation}
  A^{(n)}=iJ_nA_++iJ_{-n}A_--\delta_{n,0}M_0,
\end{equation}
where
\begin{equation}
  M_0\equiv\mat{\frac{\kappa}{2}-i\Delta&&&\\&\frac{\gamma}{2}+i\Omega&&\\&&\frac{\kappa}{2}+i\Delta\\
  &&&\frac{\gamma}{2}-i\Omega}.
\end{equation}
This includes the case discussed in the main text \eqref{eq:A matrices}
and provides a simple recipe to couple any two blocks together
and thus to engineer new types of driving schemes.
Moreover, it is straightforward to adapt this to a different system,
once the relevant matrices $M_0,A_\pm$ have been identified.

\section{The Fourier transform of the  stationary part of the autocorrelator is the measured spectrum}
\label{app:stationaryWK}
We use the definition for the spectral density from Ref.~\cite{Clerk2010a} (see also \cite{Gea-Banacloche1990,Garces2016},
where the same definition is used, also in the context of squeezing)
\begin{equation}
	\begin{aligned}
  	  S_{A\dagg A}^{\text{power}}[\omega]&\equiv\lim_{T\to\infty}\ev{|A_T[\omega]|^2}\\
  	  &=\lim_{T\to\infty}\frac{1}{T}\int_0^T\int_0^T\dd{t}\dd{t'}e^{i\omega(t'-t)}\\&\qquad\times
  	  \sum_{n,m}e^{in\delta t'+im\delta t}\ev{A^{(n)\dag}(t')A^{(m)}(t)}\\
  	  &=\lim_{T\to\infty}\frac{1}{T}\int_0^T\dd{t}\int_{-t}^{T-t}\dd{\tau}e^{i\omega\tau}\\&\qquad\times
  	  \sum_{n,m}e^{i\delta[(n+m)t+n\tau]}\ev{A^{(n)\dag}(t+\tau)A^{(m)}(t)}.
	\end{aligned}
\end{equation}
The expectation value in the last line is in fact time-translation invariant and hence independent of $t$.
Furthermore, as $T\to\infty$, the second integral becomes $\int_{-\infty}^{\infty}$.
Therefore, the expression splits into two parts
\begin{equation}
	\begin{aligned}
  S_{A\dagg A}^{\text{power}}[\omega]&=\sum_n\left( \lim_{T\to\infty}\frac{1}{T}\int_0^T\dd{t}\sum_m e^{i\delta(n+m)t} \right)\\
  &\qquad\times\left( \int_{-\infty}^\infty\dd{\tau}e^{i\delta n\tau+i\omega\tau}
  \ev{A^{(n)\dag}(\tau)A^{(m)}(0)}\right)\\
  &=\sum_{n,m}\delta_{n,-m}f_{A\dagg A}(n,m,\omega+n\delta)=S_{A\dagg A}^{(0)}(\omega).
	\end{aligned}
\end{equation}
Where it is helpful to be more precise, we note that the visibility of rotating terms
at frequency $\omega$ will decrease as $\sinc(\omega T/2)$, where $T$ is the total measurement time.

\section{Properties of the spectrum Fourier components}
\label{app:spectrum properties}
Let $A$ be governed by a time-periodic Langevin equation.
Each of its Fourier components $A^{(n)}$ obeys a Langevin equation without explicit time-dependence.
If the system assumes a stationary state (which it does if all eigenvalues of the Langevin matrix have negative real part), 
we can write the Fourier transformed Fourier components as a linear combination of
the $N$ input operators $\{F_{i,\text{in}}(\omega)\}$
(this set contains input operators and their hermitian conjugates)
\begin{equation}
  A^{(n)}(\omega)=\sum_iK^{(n)}_i(\omega)F_{i,\text{in}}(\omega),
\end{equation}
where $\vec K^{(n)}(\omega)$ is an $N$-component vector (for each Fourier component $n$)
containing the appropriate functions.
In the convention for Fourier transforms described in the main text (\cref{eq:fourier split}),
the hermitian conjugate of this equation gives
\begin{equation}
  A^{(n)\dag}(\omega)=\sum_iK_i^{(-n)*}(-\omega)F_{i,\text{in}}\dagg(\omega).
\end{equation}
The stationary part of the spectrum is (cf. \cref{eq:fourier components})
\begin{equation}
	\begin{aligned}
  	  S_{A\dagg A}^{(0)}(\omega)&=\sum_n\int\frac{\dd{\omega'}}{2\pi}\ev{A^{(n)\dag}(\omega+n\delta)A^{(-n)}(\omega')}\\
  	  &=\sum_{n,i,j}\int\frac{\dd{\omega'}}{2\pi}
  	  K_i^{(-n)*}(-\omega-n\delta)K^{(-n)}_j(\omega')\\
  	  &\qquad\times\ev{F_{i,\text{in}}\dagg(\omega+n\delta)F_{j,\text{in}}(\omega')}\\
  	  &=\sum_{n,i}n_iK_i^{(-n)*}(-\omega-n\delta)K_i^{(-n)}(-\omega-n\delta)\\
  	  &=\sum_{n,i}n_i\left|K_i^{(-n)}(-\omega-n\delta)\right|^2,
	\end{aligned}
\end{equation}
where we had to assume the noise correlators
\begin{equation}
  \ev{F_{i,\text{in}}\dagg(\omega)F_{j,\text{in}}(\omega')}=2\pi n_i\delta_{ij}\delta(\omega+\omega'),
  \label{eq:noise correlators}
\end{equation}
with thermal occupations $n_i\geq0$.
Thus the stationary part is real and positive.

Another property is $[S_{A\dagg B}^{(n)}(\omega)]\dagg=S_{B\dagg A}^{(-n)}(\omega+n\delta)$.
The proof is by expansion
\begin{equation}
	\begin{aligned}
  	  &\left[ S_{A\dagg B}^{(n)}(\omega) \right]^\dag\\
  	  &=\left[ \sum_m\int\frac{\dd{\omega'}}{2\pi}\ev{A^{(m)\dag}(\omega+m\delta)B^{(n-m)}(\omega')} \right]^\dag\\
  	  &=\sum_m\int\frac{\dd{\omega'}}{2\pi}\ev{B^{(m-n)\dag}(-\omega')A^{(-m)}(-\omega-m\delta)}\\
  	  &=\sum_m\int\frac{\dd{\omega'}}{2\pi}2\pi\delta(-\omega-m\delta-\omega')\\
  	  &\qquad\times f_{B\dagg A}(m-n,-m,-\omega-m\delta)\\
  	  &=\sum_m\int\frac{\dd{\omega'}}{2\pi}2\pi\delta(\omega'+\omega+(m+n)\delta)\\
  	  & \qquad\times f_{B\dagg A}(m,-m-n,\omega')\\
  	  &=\sum_m\int\frac{\dd{\omega'}}{2\pi}\ev{B^{(m)\dag}(\omega+(m+n)\delta)A^{(-m-n)}(\omega')}\\
  	  &=S_{B\dagg A}^{(-n)}(\omega+n\delta),
	\end{aligned}
\end{equation}
where for convenience we have again used the shorthand~\eqref{eq:shorthand}
\begin{equation}
  f_{A\dagg B}(n,m,\omega)
  \equiv\int\frac{\dd{\omega'}}{2\pi}\ev{A^{(n)\dag}(\omega)B^{(m)}(\omega')},
\end{equation}
which assumes noise correlators of the form~\eqref{eq:noise correlators}.

\section{Full solution to bichromatically driven optomechanical system in RWA}\label{app:full solution}
In RWA, the infinite set of differential equations~\eqref{eq:floquet eom} 
decouples into sets of four. 
The blocks disconnected from input operators will decay and vanish in the steady state.
Thus only two blocks (mutually hermitian conjugates) are non-zero.
The problem reduces to solving
\begin{multline}
  \mat{\chi_c^{-1}(\omega)&-iG_-&0&-iG_+\\
  -iG_-&\chi_m^{-1}(\omega)&-iG_+&0\\
  0&iG_+&\chi_c^{-1*}(-\omega+\delta)&iG_-\\
  iG_+&0&iG_-&\chi_m^{-1*}(-\omega+\delta)}\\\times
  \mat{d^{(0)}(\omega)\\b^{(0)}(\omega)\\d^{(1)\dag}(\omega)\\b^{(1)\dag}(\omega)}
  =\mat{\sqrt{\kappa}d_{\text{in}}(\omega)\\\sqrt{\gamma}b_{\text{in}}(\omega)\\0\\0},
  \label{eq:4x4 RWA system}
\end{multline}
with the cavity and mechanical response functions $\chi_c^{-1}(\omega)=\kappa/2-i(\omega+\Delta)$ and $\chi_m^{-1}(\omega)=\gamma/2-i(\omega-\Omega)$, respectively.

Eliminating the light field we find
\begin{multline}
  \mat{\chi_m^{-1}(\omega)-i\Sigma_{00}(\omega)&-i\Sigma_{01}(\omega)\\i\Sigma_{01}^*(-\omega+\delta)
  &\chi_m^{-1*}(-\omega+\delta)+i\Sigma_{00}^*(-\omega+\delta)}\\\times
  \mat{b^{(0)}\\b^{(1)\dag}}
  =\mat{\sqrt{\gamma}&iG_-\sqrt{\kappa}\chi_c(\omega)\\0&-iG_+\sqrt{\kappa}\chi_c(\omega)}
  \mat{b_{\text{in}}\\d_{\text{in}}},
  \label{eq:b solution}
\end{multline}
with
\begin{equation}
  \begin{aligned}
  	\Sigma_{00}(\omega)&=i\left[ G_-^2\chi_c(\omega)-G_+^2\chi_c^*(-\omega+\delta) \right],\\
  	\Sigma_{01}(\omega)&=iG_-G_+\left[ \chi_c(\omega)-\chi_c^*(-\omega+\delta) \right].
  \end{aligned}
\end{equation}

This allows us to write the system operators in terms of input operators
\begin{equation}
  \mat{b^{(0)}(\omega)\\b^{(1)\dag}(\omega)}=
  \mat{a(\omega)&c(\omega)\\f(\omega)&g(\omega)}\mat{b_{\text{in}}(\omega)\\d_{\text{in}}(\omega)},
\end{equation}
with
\begin{subequations}
	\begin{align}
    	a(\omega)={}& A^{-1}(\omega)\sqrt{\gamma}\left[ \chi_m^{-1*}(-\omega+\delta)+i\Sigma_{00}^*(-\omega+\delta) \right],\\
  		c(\omega)={}&iA^{-1}(\omega)\sqrt{\kappa}\chi_c(\omega)G_-\\&
  		\quad\times\left[ \chi_m^{-1*}(-\omega+\delta)+\G^2\chi_c^*(-\omega+\delta) \right],\nonumber\\
		f(\omega)={}&-iA^{-1}(\omega)\sqrt{\gamma}\Sigma_{01}(\omega),\\
  		g(\omega)={}&-iA^{-1}(\omega)\sqrt{\kappa}\chi_c(\omega)G_+\\&
  		\quad\times\left[\chi_m^{-1}(\omega)+\G^2\chi_c^*(-\omega+\delta) \right],\nonumber
	\end{align}
  	  	  \label{eq:aux fun}
\end{subequations}
where $\G^2\equiv G_-^2-G_+^2$ and
$A(\omega)$ is the determinant of the matrix on the left hand side of~\cref{eq:b solution},
\begin{multline}
  A(\omega)=\left[ \chi_m^{-1*}(-\omega+\delta)+i\Sigma_{00}^*(-\omega+\delta)\right]\\\times
  \left[ \chi_m^{-1}(\omega)-i\Sigma_{00}(\omega) \right]
  -\Sigma_{01}(\omega)\Sigma_{01}^*(-\omega+\delta).
\end{multline}

The analytical solution can be used to find spectrum Fourier components,
employing~\cref{eq:fourier components},
\begin{subequations}
	\begin{align}
		\begin{split}
  			S_{b\dagg b}^{(0)}(\omega)&=|a(-\omega)|^2n_{\text{th}}\\
  			&\quad+|f(\omega+\delta)|^2(n_{\text{th}}+1)
  			+|g(\omega+\delta)|^2,
		\end{split}\\
  		\begin{split}
			S_{b b\dagg}^{(0)}(\omega)&=(n_{\text{th}}+1)|a(\omega)|^2+|c(\omega)|^2\\
			&\qquad+n_{\text{th}}|f(-\omega+\delta)|^2,
  		\end{split}\\
  		\begin{split}
  			S_{xx}^{(0)}(\omega)&=(n_{\text{th}}+1)\left( |a(\omega)|^2+|f(\omega+\delta)|^2\right)+|c(\omega)|^2\\
  			&\,+n_{\text{th}}\left( |a(-\omega)|^2+|f(-\omega+\delta|^2 \right)+|g(\omega+\delta)|^2,
  		\end{split}\\
  		\begin{split}
			S_{b b}^{(-1)}(\omega)&=(n_{\text{th}}+1)a(\omega)f^*(\omega)+c(\omega)g^*(\omega)\\
			&\qquad+n_{\text{th}}f^*(-\omega+\delta)a(-\omega+\delta)\\
			&=[S_{b\dagg b\dagg}^{(1)}(\omega-\delta)]\dagg.
  		\end{split}
	\end{align}
\end{subequations}

An important special case~\cite{Pirkkalainen2015,Lecocq2015} is the symmetric detuning $\delta=2\Omega+\eps$, $\Delta=-\Omega-\eps/2=-\delta/2$.
Crucially, this leads to $\chi_c^*(-\omega+\delta)=\chi_c(\omega)$, which implies
\begin{equation}
  \Sigma_{00}=i\chi_c(\omega)\G^2,\qquad \Sigma_{01}=0.
\end{equation}
Thus, the determinant $A(\omega)$ takes a particularly simple form
\begin{equation}
  A(\omega)=\left[ \chi_m^{-1}(\omega-\eps)+\chi_c(\omega)\G^2 \right]
  \left[ \chi_m^{-1}(\omega)+\chi_c(\omega)\G^2 \right]
\end{equation}
and so do the auxiliary functions
\begin{equation}
  \begin{aligned}
	a(\omega)&=\sqrt{\gamma}/\left[ \chi_m^{-1}(\omega)+\chi_c(\omega)\G^2 \right],\\
	c(\omega)&=i\sqrt{\kappa}G_-\chi_c(\omega)/\left[ \chi_m^{-1}(\omega)+\chi_c(\omega)\G^2 \right],\\
	f(\omega)&=0,\\
	g(\omega)&=-i\sqrt{\kappa}\chi_c(\omega)G_+/\left[ \chi_m^{-1}(\omega-\eps)+\G^2\chi_c(\omega) \right].
  \end{aligned}
\end{equation}
And the spectra are
\begin{subequations}
	\begin{align}
		\begin{split}
  			S_{xx}^{(0)}(\omega)&=\frac{(n_{\text{th}}+1)\gamma+\kappa G_-^2|\chi_c(\omega)|^2}{|\chi_m^{-1}(\omega)+\chi_c(\omega)\G^2|^2}\\
  			&\quad+\frac{\gamma n_{\text{th}}+\kappa G_+^2|\chi_c(-\omega)|^2}{|\chi_m^{-1}(-\omega)+\chi_c(-\omega)\G^2|^2},
  			\label{eq:lab frame mech}
		\end{split}\\
  		S_{xx}^{(1)}(\omega)&=\frac{\kappa G_-G_+\chi_c(-\omega)}{\sigma^*(-\omega)\left[ \chi^{-1}_m(-\omega-\eps)+\chi_c(-\omega)\G^2 \right]}\\
  		&=[S_{xx}^{(-1)}(-\omega)]^*,\nonumber
	\end{align}
\end{subequations}
where we have introduced $\sigma(\omega)=\G^2+\chi_m^{-1}(\omega)\chi_c^{-1}(\omega)$.

We can employ \cref{eq:rot frame} for the physical spectrum in the special rotating frame.
It has two parts.
One is the previously stationary part,
which corresponds to the radially symmetric contribution to the Wigner density
(and therefore it remains stationary, despite going into a rotating frame)
\begin{equation}
	\begin{aligned}
  	  U(\omega)&=S_{b\dagg b}^{(0)}(\omega-\delta/2)+S_{b b\dagg}^{(0)}(\omega+\delta/2)\\
  	  &=\kappa\left[ \frac{G_-^2}{|\sigma(\omega+\delta/2)|^2}+\frac{G_+^2}{|\sigma(-\omega+\delta/2)|^2} \right]\\
  	  &+\frac{\gamma}{|\chi_c(\omega+\delta/2)|^2}
  	  \left[ \frac{n_{\text{th}}+1}{|\sigma(\omega+\delta/2)|^2}
  	  +\frac{n_{\text{th}}}{|\sigma(-\omega+\delta/2)|^2} \right]
	\end{aligned}
\end{equation}
The other one stems from the previously rotating parts
\begin{equation}
	\begin{aligned}
		V(\omega)&=S_{b\dagg b\dagg}^{(1)}(\omega-\delta/2)+S_{b b}^{(-1)}(\omega+\delta/2)\\
		&=-2\kappa G_-G_+\\
		&\times\Re\left\{ \frac{1}{|\sigma(\omega+\delta/2)|^2-i\eps\chi_c^{-1*}(\omega+\delta/2)\sigma(\omega+\delta/2)} \right\}.
	\end{aligned}
\end{equation}
Finally,
\begin{equation}
  S_{X_\mp X_\mp}(\omega)=U(\omega)\pm V(\omega).
  \label{eq:symmetric rot spectrum}
\end{equation}

\section{Weak-coupling approximation to a bichromatically driven optomechanical system}\label{app:weak}
Our approach in this section will be to perturb around the mechanical spectrum in the absence of coupling.
We will do so up to second order in $G_\pm$.

The equations of motion in RWA~\eqref{eq:RWA langevin}, split up into Fourier components, are
\begin{subequations}
	\begin{align}
		\begin{split}
			\dot d^{(n)}&=\left( -in\delta+i\Delta-\frac{\kappa}{2} \right)d^{(n)}+\sqrt{\kappa}\delta_{n,0}d_{\text{in}}\\
			&\qquad+i\left( G_+b^{(n+1)\dag}+G_-b^{(n)} \right),
		\end{split}\\
		\begin{split}
			\dot b^{(n)}&=\left( -in\delta-i\Omega-\frac{\gamma}{2} \right)b^{(n)}+\sqrt{\gamma}\delta_{n,0}b_{\text{in}}\\
			&\qquad+i\left( G_+d^{(n+1)\dag}+G_-d^{(n)} \right).
		\end{split}
	\end{align}
\end{subequations}
The unperturbed mechanical spectrum consists only of $b^{(0)}$ and $b^{(0)\dag}$.
Thus, to first order,
\begin{subequations}
	\begin{align}
		d^{(0)}(\omega)&=\chi_c(\omega)\left( \sqrt{\kappa}d_{\text{in}}(\omega)+iG_-b^{(0)}(\omega)\right),\\
		d^{(-1)}(\omega)&=\chi_c(\omega+\delta)iG_+b^{(0)\dag}(\omega).
	\end{align}
\end{subequations}
We can now determine $b^{(0)}$ without knowledge of $b^{(-1)}$
\begin{multline}
  \left( -i\omega+i\Omega+\frac{\gamma}{2}-G_+^2\chi_c^*(-\omega+\delta)+G_-^2\chi_c(\omega) \right)b^{(0)}(\omega)\\
  =\sqrt{\gamma}b_{\text{in}}(\omega)+i\sqrt{\kappa}G_-\chi_c(\omega)d_{\text{in}}(\omega)+\O(G_\pm^3).
  \label{eq:weak b0 eom}
\end{multline}
The reason for this is that $b^{(-1)}=\O(G_\pm)$, such that the effect $b^{(-1)}$ has on $b^{(0)}$
(via the optical field) is at least $\O(G_\pm^3)$.
The LHS of \eqref{eq:weak b0 eom} is the modified response function
\begin{equation}
  \tilde\chi_m(\omega)=\left[\frac{\tilde \gamma(\omega)}{2}-i(\omega-\tilde\Omega(\omega))\right]^{-1}
  \label{eq:mod mech response fun}
\end{equation}
with
\begin{subequations}
	\begin{align}
		\tilde\gamma(\omega)&=\gamma+\kappa\left( |\chi_c(\omega)|^2G_-^2-|\chi_c(-\omega+\delta)|^2G_+^2 \right),
  	  	  \label{eq:gammatildefull}\\
  	  	  \begin{split}
			\tilde\Omega(\omega)&=\Omega+|\chi_c(\omega)|^2(\omega+\Delta)G_-^2\\
			&\qquad+|\chi_c(-\omega+\delta)|^2(-\omega+\delta+\Delta)G_+^2.
  	  	  \end{split}
  	  	  \label{eq:Omegatildefull}
	\end{align}
\end{subequations}
The mechanical response function \eqref{eq:mod mech response fun} strongly suppresses contributions away from $\omega=-\tilde\Omega\approx-\Omega$.
In comparison to $\chi_m$, $\chi_c$ is flat (if $\tilde\gamma\ll\kappa$),
such that we can approximate $\chi_c(\omega)\approx\chi_c(-\Omega)$.
For $\Delta=-\Omega$ the corrections simplify to Eqs.~(\ref{eq:gammatilde}), and
\begin{equation}
  b^{(0)}(\omega)=\tilde\chi_m(\omega)\sqrt{\gamma}b_{\text{in}}
  +\frac{2iG_-}{\sqrt{\kappa}}\chi_m(\omega)d_{\text{in}}.
  \label{eq:weak b0}
\end{equation}

We would like the same accuracy for the rotating components, so we keep the next order in $d$
\begin{equation}
  \begin{aligned}
	d^{ (0)}(\omega)&= \chi_c(\omega)\left(\sqrt{\kappa}d_{\text{in}}(\omega)+iG_-b^{(0)}+iG_+b^{(1)\dag}(\omega)\right),\\
	d^{(-1)}(\omega)&=i\chi_c(\omega+\delta)\left(G_+b^{(0)\dag}(\omega)+G_-b^{(-1)}(\omega)\right).
  \end{aligned}
  \label{eq:eq for d}
\end{equation}
Then
\begin{equation}
  \chi_m^{-1*}(-\omega+\delta)b^{(1)\dag}(\omega)=
  -i\left(G_+d^{(0)}(\omega)+G_-d^{(1)\dag}(\omega) \right).
\end{equation}
We substitute for $d$ with \cref{eq:eq for d}
\begin{multline}
  b^{(1)\dag}(\omega)
  =\chi_m^{\prime*}(-\omega+\delta)\left[ -iG_+\chi_c(\omega)\sqrt{\kappa}d_{\text{in}}(\omega)\right.\\
  \left.+G_+G_-\left( \chi_c(\omega)-\chi_c^*(-\omega+\delta) \right)b^{(0)}(\omega)\right],
\end{multline}
where $b^{(0)}=\chi_m(\omega)\sqrt{\gamma}b_{\text{in}}$ in the absence of driving 
(the second order corrections to $b^{(0)}$ would be fourth order in this equation).
With a differently modified response function $\chi_m'=[\gamma'/2-i(\omega-\Omega')]^{-1}$, with
\begin{subequations}
	\begin{align}
		\gamma'(\omega)&=\gamma+\kappa\left(|\chi_c(-\omega+\delta)|^2G_-^2-|\chi_c(\omega)|^2G_+^2\right),\\
		\begin{split}
			\Omega'(\omega)&=\Omega+|\chi_c(\omega)|^2(\omega+\Delta)G_+^2\\
			&\qquad+|\chi_c(-\omega+\delta)|^2(-\omega+\delta+\Delta)G_-^2.
		\end{split}
	\end{align}
\end{subequations}
Comparing with Eq.~(\ref{eq:gammatildefull}), we see that the corrections have the same form,
but with the frequencies interchanged.
The reason that the picture is reversed is that
$b^{(1)\dag}$ rotates in sync with the upper drive and not with the lower one as $b^{(0)}$ does.
In the case $\Delta=-\Omega,\omega=\Omega$, they are mirrored versions of Eq.~(\ref{eq:gammatildefull})
\begin{subequations}
	\begin{align}
		\gamma'&=\gamma+\frac{4}{\kappa}\left( \frac{G_-^2}{1+4\eps^2/\kappa^2}-G_+^2 \right),\\
		\Omega'&=\Omega+\frac{\eps G_-^2}{\kappa^2/4+\eps^2}.
	\end{align}
\end{subequations}

We can neglect the second-order perturbation on the first order quantities $b^{(1)\dag},b^{(-1)}$,
because they appear to third order on the level of spectrum calculations,
such that
\begin{equation}
  b^{(1)\dag}(\omega)=-\chi_m^*(-\omega+\delta)\frac{2iG_+}{\sqrt{\kappa}}d_{\text{in}}.
  \label{eq:weak b1}
\end{equation}
In the main text we use the modified parameters $\tilde\gamma,\tilde\Omega$ in \eqref{eq:weakeom2}.
With this replacement, \cref{eq:weak b0,eq:weak b1} yield \cref{eq:weak aux funs}
It might seems surprising to use $\tilde\gamma,\tilde\Omega$ instead of $\gamma',\Omega'$,
but is allowed, as the corrections are third order.
We mainly do that for convenience, because it makes the subsequent analysis more transparent.
Comparing to the full solution and looking at \cref{fig:squeezingloss}, we see that our approximation is reasonable.
In fact, we cannot use $\gamma'$, because it crosses zero for relatively small detunings $\eps<\kappa$ when $\cc>n_{\text{th}}$,
which leads to a divergence. 

In order to derive a master equation, we define $\tilde b\equiv e^{i\delta t/2}b$, and assume $\delta=2\Omega$.
Then $\beta\equiv(G_-\tilde b+G_+\tilde b\dagg)/\G$ obeys
\begin{equation}
  \dot\beta=\left( -\frac{\gamma}{2}-\frac{2\G^2}{\kappa} \right)\beta+\sqrt{\gamma}\beta_{\text{in}}
  +\frac{2i\G}{\sqrt{\kappa}}\tilde d_{\text{in}},
\end{equation}
where $\tilde d_{\text{in}}\equiv e^{i\Omega t}d_{\text{in}}$.
The associated quantum master equation is (NB in frame rotating with the mechanical frequency $\Omega$)
\begin{equation}
  \dot\dens=\left(\gamma n_{\text{th}}\D[\tilde b\dagg]+\gamma(n_{\text{th}}+1)\D[\tilde b]
  +\frac{4\G^2}{\kappa}\D[\beta]\right)\dens.
\end{equation}
This agrees with Ref.~\cite{Kronwald2013}.
The physics here is that the drives cool the Bogoliubov mode $\beta$ close to its ground state, which is a squeezed state for the rotating quadrature $\tilde b+\tilde b\dagg$~\cite{Kronwald2013}.

\section{Analysis of instability within RWA}\label{app:stability analysis}
\begin{figure*}[t]
  \includegraphics[width=\linewidth]{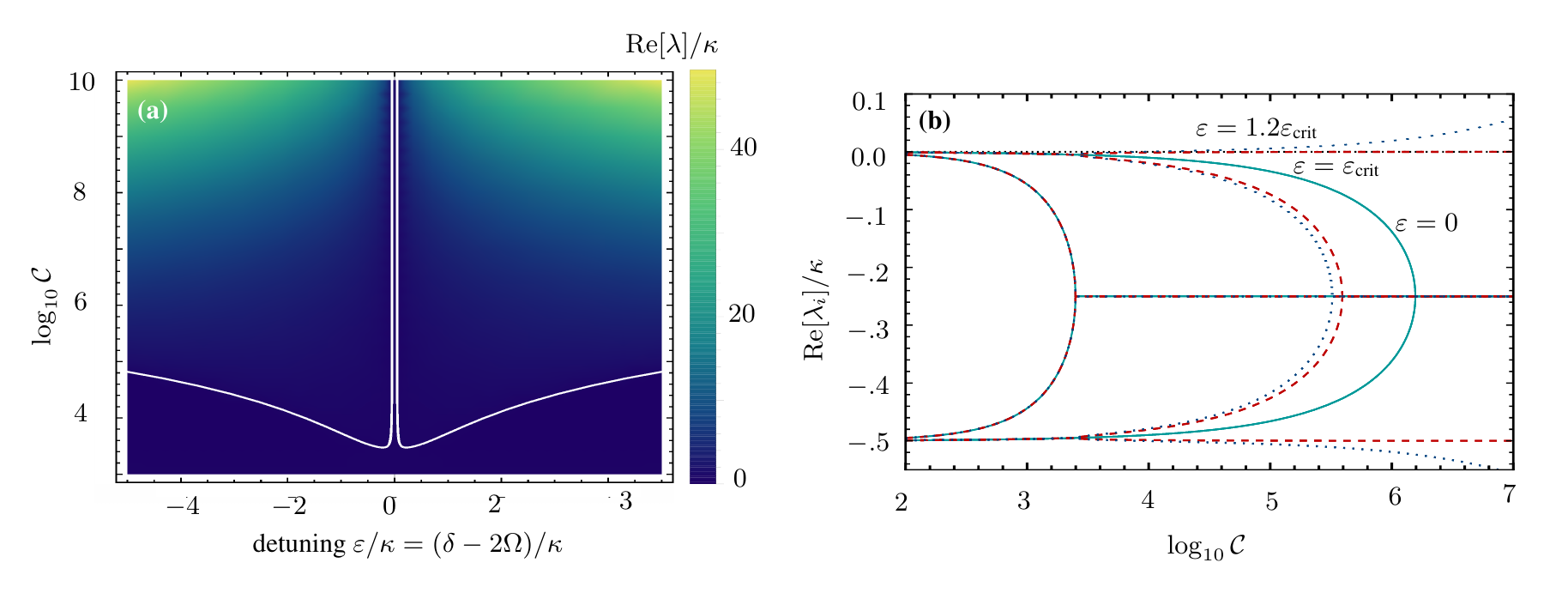}
  \caption{\textbf{Analysis of instability within RWA.}
  	Parameters are $\gamma/\kappa=10^{-4},n_{\text{th}}=10,\cc=2\times 10^3,$ and $\Delta=-\Omega$.
  	$\Omega/\kappa$ is irrelevant in RWA.
	\textbf{(a) Boundary of stability.} The white curve is the analytical result for the
  	boundary of stability (\ref{eq:stability bounds}).
	The color scale gives the real part of the eigenvalue with the largest real part $\lambda$ of the matrix (\ref{eq:4x4 RWA system}).
	As $\cc\to\infty$, a ``stability corridor" remains,
	$\eps_{\text{crit}}=\pm\sqrt{\kappa\gamma(1+2n_{\text{th}})}$.
	The corridor collapses without RWA.
	\textbf{(b) Eigenvalues as a function of cooperativity.}
	Real part of the eigenvalues of (\ref{eq:4x4 RWA system}) as a function of  cooperativity $\cc$ for optimal driving \cref{eq:optimal driving strength},
	and detuning $\eps=0,\eps_{\text{crit}},1.2\eps_{\text{crit}}$,
	in dark blue (solid), red (dashed) and turquoise (dotted).
	In the strictly stable regime all eigenvalues converge to have the same real part $(\kappa+\gamma)/4$
	at large cooperativities $\cc$.
	At the critical detuning two eigenvalues remain at $\gamma/2$ and two at $\kappa/2$ for all $\cc$. 
	Above the critical detuning, there exists a value of $\cc$ above which the system is unstable.
  }
  \label{fig:stability}
\end{figure*}
To study the instability we employ the Routh-Hurwitz criterion, according to which a system is unstable if the matrix $M$ in $\dot{\vec x}=M\vec x$ has an eigenvalue with positive real part.

Let us call the matrix on the LHS of \eqref{eq:4x4 RWA system} $K(\omega)$. In our case, $K(0)=-M$. Thus we can write $K(\omega)=-M-i\omega I_4$, where $I_4$ is the $4\times4$ identity matrix. The eigenvalues of $K(\omega)$ satisfy the secular equation
\begin{equation}
  \det[-M-(i\omega+\lambda)I_4]=0.
\end{equation}
Thus, if $\lambda$ is an eigenvalue of $M$, then $-\lambda+i\omega$ is an eigenvalue of $K(\omega)$.
We conclude that if $\Re[\lambda]=0$, $K(\Im[\lambda])$ is singular, and \emph{vice versa},
which marks the onset of instability.

Assuming $\Delta=-\Omega$, it turns out that $\det[K(\Omega+\eps/2)]$ is purely real and
\begin{equation}
  \det[K(\Omega+\eps/2)]=\sigma(\omega)\sigma(\omega-\eps)-\eps^2G_+^2,
  \label{eq:square bracket}
\end{equation}
with $\sigma(\omega)=\G^2+\chi_m^{-1}(\omega)\chi_c^{-1}(\omega)$ and $\delta=2\Omega+\eps$.
Its imaginary part is zero at $\omega=\Omega+\eps/2$, so we are left with
\begin{equation}
  0=\left( \G^2+\frac{\gamma\kappa}{4}-\frac{\eps^2}{4} \right)^2+\frac{\eps^2(\kappa+\gamma)^2}{16}-\eps^2 G_+^2,
\end{equation}
which gives
\begin{multline}
  \eps_\pm^2=4(G_-^2+G_+^2)-\frac{\kappa^2+\gamma^2}{2}\\
  \pm\sqrt{\left[ \frac{\kappa^2+\gamma^2}{2}-4(G_-^2+G_+^2) \right]^2-\left(4\G^2+\gamma\kappa\right)^2}.
  \label{eq:stability bounds}
\end{multline}
$\eps_\pm$ is complex if the term under the root is negative, i.e., if
\begin{equation}
  \cc\leq\left( \frac{\kappa+\gamma}{2\sqrt{\kappa\gamma}}+\sqrt{1+2n_{\text{th}}} \right)^2.
  \label{eq:total stability}
\end{equation}
In \cref{eq:total stability} we have used the optimal driving strengths, see \cref{eq:optimal driving strength} or Ref.~\cite{Kronwald2013}.
We conclude that there is an instability for $\eps_-<|\eps|<\eps_+$.
Note that the stability regions are symmetric in $\eps$ with stability at $\eps=0$.
Because of condition~\eqref{eq:total stability},
we can only study large detuning for small cooperativities.
As $\cc\to\infty$, $\eps_{-}\to\eps_{\text{crit}}=\pm\sqrt{\kappa\gamma(1+2n_{\text{th}})}$,
so there is a ``stability corridor'' in between $\pm\eps_-$ even at largest cooperativities,
which is shown in \cref{fig:stability}.
Once we numerically include counterrotating terms, the stability corridor is lost.
Note that we have assumed $\Delta=-\Omega$
and that if $G_+>G_-$ the system may be unstable for all detunings $\delta$.

\section{The optical spectrum $S_{d\dagg d}$}\label{app:sdd spectrum}
Following the same steps as in the main text, we can write the optical system operators in terms of the input operators
\begin{equation}
  \mat{d^{(0)}(\omega)\\d^{(1)\dag}(\omega)}=
  \mat{\tilde a(\omega)&\tilde c(\omega)\\\tilde f(\omega)&\tilde g(\omega)}
  \mat{d_{\text{in}}(\omega)\\b_{\text{in}}(\omega)}.
\end{equation}
We obtain the functions in the matrix by a calculation analogous to the one in \cref{app:full solution}.
Because of the symmetry of the equations of motion in the RWA, this amounts to swapping $\gamma\leftrightarrow\kappa,
\chi_c\leftrightarrow\chi_m,b_{\text{in}}\leftrightarrow d_{\text{in}}.$
Thus,
\begin{equation}
  \begin{aligned}
  	\tilde a(\omega)&=\tilde A^{-1}(\omega)\sqrt{\kappa}
    \left[\chi_c^{-1*}(-\omega+\delta)+i\tilde \Sigma_{00}^*(-\omega+\delta) \right],\\
  	\tilde c(\omega)&=i\tilde A^{-1}(\omega)\sqrt{\gamma}\chi_m(\omega)G_-\\
  	&\qquad\times\left[ \chi_c^{-1*}(-\omega+\delta)+\G^2\chi_m^*(-\omega+\delta) \right],\\
	\tilde f(\omega)&=-i\tilde A^{-1}(\omega)\sqrt{\kappa}\tilde \Sigma_{01}(\omega),\\
  	\tilde g(\omega)&=-i\tilde A^{-1}(\omega)\sqrt{\gamma}\chi_m(\omega)G_+\\
  	&\qquad\times\left[\chi_c^{-1}(\omega)+\G^2\chi_m^*(-\omega+\delta) \right],
  \end{aligned}
  \label{eq:prime aux fun}
\end{equation}
with 
\begin{multline}
  \tilde A(\omega)=\left[ \chi_c^{-1*}(-\omega+\delta)+i\tilde \Sigma_{00}^*(-\omega+\delta)\right]\\\times
  \left[ \chi_c^{-1}(\omega)-i\tilde \Sigma_{00}(\omega) \right]
  -\tilde\Sigma_{01}(\omega)\tilde \Sigma_{01}^*(-\omega+\delta),
\end{multline}
and
\begin{equation}
  \begin{aligned}
  	\tilde \Sigma_{00}(\omega)&=i\left[ G_-^2\chi_m(\omega)-G_+^2\chi_m^*(-\omega+\delta) \right],\\
  	\tilde \Sigma_{01}(\omega)&=iG_-G_+\left[ \chi_m(\omega)-\chi_m^*(-\omega+\delta) \right].
  \end{aligned}
\end{equation}
The stationary part of the optical spectrum is
\begin{equation}
  S_{d\dagg d}^{(0)}(\omega)=|\tilde c(-\omega)|^2n_{\text{th}}
  +|\tilde f(\omega+\delta)|^2+|\tilde g(\omega+\delta)|^2(n_{\text{th}}+1).
\end{equation}

Note that the same spectra can be obtained by employing the formulae in \cref{sec:readout}
for the case $\delta=\delta_2$.
As the mechanical spectrum one has to use the one derived in \cref{app:full solution}, in particular \cref{eq:special case d2 spectrum}.
Further note that the output spectrum is trivially related to $S_{d\dagg d}$, since the input-output relation in our case is
\begin{equation}
	d_{\text{out}}=d_{\text{in}}-\sqrt{\kappa}d,
	\label{eq:input-output}
\end{equation}
such that
\begin{equation}
	S_{d_{\text{out}}\dagg d_{\text{out}}}^{(n)}(\omega)=\kappa S_{d\dagg d}^{(n)}(\omega).
	\label{eq:output spectrum}
\end{equation}

\section{Readout spectra in second mode}\label{app:second mode}
In this section we provide more details on the calculation of the readout spectra.
We split \cref{eq:second mode langevin} into Fourier components
\begin{multline}
  \sum_{n,m}e^{in\delta t+im\delta_2t}\left[ (in\delta+im\delta_2)d_2^{(n,m)}+\dot d_2^{(n,m)} \right]\\
  =\sum_{n,m}e^{in\delta t+im\delta_2t}\left[ \left( i\Delta_2-\frac{\kappa_2}{2} \right)d_2^{(n,m)}
  	+\sqrt{\kappa_2}d_{2,\text{in}}\delta_{n,0}\delta_{m,0}\right.\\
  +\left.i\left( G_{2+}x^{(n,m+1)}+G_{2-}x^{(n,m)} \right)\right].
\end{multline}

If $\delta\neq\delta_2$ (and they do not have a common multiple), we have
$b^{(n,m\neq0)}=0$. Thus, with $x^{(n,m)}\equiv b^{(n,m)}+b^{(n,m)\dag}$,
\begin{subequations}
	\begin{align}
  		d_2^{(n,0)}&=\chi_{2}(\omega-n\delta)\left[ \sqrt{\kappa_2}\delta_{n,0}d_{2,\text{in}}+iG_{2-}x^{(n,0)}\right]\\
  		d_2^{(n,0)\dag}&=\chi_{2}^*(-\omega+n\delta)
  		\left[\sqrt{\kappa_2}\delta_{n,0}d_{2,\text{in}}\dagg-iG_{2-}x^{(n,0)}\right],\\
  		d_2^{(n,-1)}&=\chi_2(\omega-n\delta+\delta_2)iG_{2+}x^{(n,0)},\\
  		d_2^{(n,1)\dag}&=-\chi_2^*(-\omega+n\delta+\delta_2)iG_{2+}x^{(n,0)}.
	\end{align}
\end{subequations}
A substitution into \cref{eq:fourier components} yields \eqref{eq:d2 stationary part}.

For the special choice $\delta_2=\delta$,
\begin{subequations}
	\begin{align}
		\begin{split}
  			d_2^{(n)}(\omega)&=\chi_2(\omega-n\delta)\\
  			&\left[ \sqrt{\kappa_2}\delta_{n,0}d_{2,\text{in}}
  				+i\left( G_{2+}x^{(n+1)}+G_{2-}x^{(n)} \right)\right],
		\end{split}\\
		\begin{split}
  			d_2^{(n)\dag}(\omega)&=\chi_2^*(-\omega+n\delta)\\
  			&\left[ \sqrt{\kappa_2}\delta_{n,0}d_{2,\text{in}}\dagg
  				-i\left( G_{2+}x^{(n-1)}+G_{2-}x^{(n)} \right)\right].
		\end{split}
	\end{align}
\end{subequations}
Again we substitute into \cref{eq:fourier components} to get \eqref{eq:d2 xx spectrum}.

\bibliographystyle{apsrev4-1.bst}

\end{document}